\documentclass[11pt]{article}

\usepackage{mathrsfs}
\usepackage{appendix}
\usepackage{natbib}
\usepackage{extarrows}
\usepackage{graphicx,setspace,lscape,longtable,epstopdf,xr}
\usepackage{mathrsfs,amsmath,amsthm,amssymb,color}
\usepackage{epsfig,graphicx,pdfpages}
\usepackage{rotating}
\usepackage{float}
\usepackage{bm}
\usepackage{ragged2e}
\usepackage{todonotes}
\usepackage{hyperref}
\hypersetup{
	colorlinks=true,
	linkcolor=blue,
	citecolor=blue,
	urlcolor=blue}
\usepackage{booktabs}
\usepackage{algorithm}
\usepackage{algpseudocode}
\usepackage{amsmath}
\usepackage{graphicx}
\usepackage{subfigure}
\usepackage{hyperref}
\def\be{\begin{eqnarray}}
\def\ee{\end{eqnarray}}
\bibpunct{(}{)}{;}{a}{,}{,}

\newtheorem{theorem}{Theorem}
\newtheorem{lemma}{Lemma}
\newtheorem{proposition}{Proposition}

\newtheorem{assumption}{Condition}
\def\ve{\varepsilon}

\def\tr{\mbox{tr}}

\def\beq{\begin{equation}}
\def\eeq{\end{equation}}
\def\beqr{\begin{eqnarray}}
\def\eeqr{\end{eqnarray}}
\def\beqrs{\begin{eqnarray*}}
	\def\eeqrs{\end{eqnarray*}}
\def\bet{\begin{theorem}}
	\def\eet{\end{theorem}}
\def\bel{\begin{lemma}}
	\def\eel{\end{lemma}}
\def\bep{\begin{proposition}}
	\def\eep{\end{proposition}}
\def\bg{\begin{figure}[tbph]\begin{center}}
		\def\eg{\end{center}\end{figure}}

\def\bc{\begin{center}}
	\def\ec{\end{center}}

\def\y{\mathbf{y}}
\def\B{\mathbf{B}}
\newtheorem{remark}{Remark}

\def\wt{\widetilde}

\def\wh{\widehat}

\def\ol{\overline }

\def\bB{\mathbf{B}}

\def\A{\mathbf A}
\def\mC{\mathcal C}
\def\mN{\mathcal{N}}

\def\mG{\mathbb G}
\def\mR{\mathbb{R}}

\def\mS{\mathbb S}

\def\bY{\bm Y}
\def\mS{\mathcal S}

\def\cX{\mathcal{X}}

\def\bx{\mathbf{x}}
\def\bz{\mathbf{z}}

\def\bv{\mathbf{v}}
\def\bb{\mathbf{b}}

\def\argmin{\mbox{argmin}}
\DeclareMathOperator*{\argmax}{argmax} 

\newcommand{\bzeta}{\boldsymbol{\zeta}}

\newcommand{\bbeta}{\boldsymbol{\beta}}
\newcommand{\bxi}{\boldsymbol{\xi}}
\def\btheta{\boldsymbol{\theta}}
\def\bve{\boldsymbol{\varepsilon}}
\def\bSigma{\boldsymbol{\Sigma}}

\def\bPhi{\boldsymbol{\Phi}}

\def\bnu{\boldsymbol{\nu}}
\def\bmu{\boldsymbol{\mu}}

\def\bvarphi{\boldsymbol{\varphi}}
\newcommand{\RNum}[1]{\uppercase\expandafter{\romannumeral #1\relax}}
\def\bX{\mathbf{X}}

\def\by{\mathbf{y}}

\def\M{\mathbf{M}}

\def\bg{\mbox{\boldmath $g$}}
\def\bI{\mbox{\boldmath $I$}}

\def\se{\mbox{SE}}

\def\boxit#1{\vbox{\hrule\hbox{\vrule\kern6pt\vbox{\kern6pt#1\kern6pt}\kern6pt\vrule}\hrule}}

\newcommand{\xggrev}[1]{{\color{magenta}{#1}}}

\numberwithin{equation}{section}
\begin{document}
	\begin{center}
		{\bf\Large Simultaneous Estimation and Group Identification for Network Vector Autoregressive Model with Heterogeneous Nodes}\\
		\bigskip
		Xuening Zhu$^1$, Ganggang Xu$^{2}$, and Jianqing Fan$^{3,4}$

		{\it 
             $^1$Fudan University,  China;
			$^2$University of Miami, USA;\\
			$^3$Capital University of Economics and Business, China\\
            $^4$Princeton University, USA
}
		
		%

	\end{center}

\begin{footnotetext}[1]{Xuening Zhu is supported by the National Natural Science
    Foundation of China (nos. 71991470, 72222009, 71991471, 71991472).
    Xuening Zhu and Ganggang Xu are joint first authors, and Jianqing Fan is the
    corresponding author.}
\end{footnotetext}
	
	\begin{singlespace}
		\begin{abstract}
Individuals or companies in a large social or financial network often display rather heterogeneous behaviors for various reasons. In this work, we propose a  network vector autoregressive model with a latent group structure to model heterogeneous dynamic patterns observed from network nodes, for which group-wise network effects and time-invariant fixed-effects can be naturally incorporated. In our framework, the model parameters and network node memberships can be simultaneously estimated by minimizing a least-squares type objective function.
In particular, our theoretical investigation allows the number of latent groups $G$ to be over-specified when achieving the estimation consistency of the model parameters and group memberships, which significantly improves the robustness of the proposed approach. When $G$ is correctly specified, valid statistical inference can be made for model parameters based on the asymptotic normality of the estimators. A data-driven criterion is developed to consistently identify the true group number for practical use. Extensive simulation studies and two real data examples are used to demonstrate the effectiveness of the proposed methodology.


		\end{abstract}
	\end{singlespace}
	
	\newpage
	
	\section{Introduction}
	High dimensional time series harvested from large network platforms such as social networks and financial networks has become increasingly available in recent years.
	Much research interest has been  devoted to model dynamics of the associated network time series. Examples include \cite{sewell2015latent,zhu2017network,zhu2018network} and references therein.
	While abundant literature is available for network time series data, one remaining challenge is how to account for the commonly encountered nodal heterogeneity. 
	For example, in a social network, users with different  education or social-economic backgrounds may have rather different posting behaviors and may interact differently with members from other social groups.	There has been scarce work on modeling such heterogeneous network effects in the literature, including the  spatial autoregression model studied in
	\cite{dou2016generalized} and the feature screening of  network nodes proposed in
	\cite{zhu2019portal}. However, both works can only model the heterogeneous network effect on an individual node level.
	In this work, we propose a network autoregression model with a latent group structure (GNAR) for jointly modeling the time series data collected from all potentially heterogeneous network nodes.
	
	Consider a network with $N$ nodes indexed by $i = 1,\cdots, N$, whose relationships are recorded through an adjacency matrix $\A = (a_{ij})\in \{0,1\}^{N\times N}$, where $a_{ij}=1$ if the $i$th node follows the $j$th node and $0$ otherwise.  By convention, we set $a_{ii}=0$ for all $i = 1,\cdots, N$.  For the $i$th node, we observe a time series of continuous variable, denoted by  $\{Y_{it}\}_{t=0}^T$, together with
	a set of node specific covariates $\bz_i\in \mR^p$. In particular, we remark that the first entry of the vector $\bz_i$ is always $1$, which corresponds to the intercept term. To account for the network heterogeneity, we assume that the network nodes can be clustered into $G$ groups with homogenous within-group regression effect and use $g_i\in \{1,\cdots, G\}$ to denote the group membership of the $i$th node. The GNAR model can be expressed as
	\begin{align}
	Y_{it} =  \sum_{j = 1,j\ne i}^N\beta_{g_ig_j}w_{ij}Y_{j(t-1)} + \nu_{g_i} Y_{i(t-1)} +
	\bz_i^\top\bzeta_{g_i} +
	\ve_{it},\quad t=1,\cdots,T,\label{gnar}
	\end{align}
	where $w_{ij} = n_i^{-1}a_{ij}$ with  $n_i = \sum_{j=1}^N a_{ij}$ being the out-degree  of node $i$, and
	$\ve_{it}$'s are independent and identically distributed random noises with a mean $0$ and variance $\sigma^2$. All model parameters as well as the node membership $g_i$'s will be estimated.

	The key assumption of the GNAR model is that nodes from the same group, say group $g$, share similar characteristics such as the node-specific momentum effect ($\nu_{g}$) and covariate-related fixed-effect ($\bzeta_{g}$). The interactions between nodes from two groups, say $g,g'$, share the same group-level network effect parameter $\beta_{gg'}$. Such assumptions are reasonable for many popular networks such as social networks.	From the estimation point of view, the GNAR model strikes a good balance between the model flexibility and complexity.	
	In the special case with $G = 1$, the GNAR model reduces to the network vector autoregression (NAR) model proposed in \cite{zhu2017network}, which may not be flexible enough
	since it requires homogeneous  network effects, momentum effects, and fixed-effects.
	In the other extreme case with $G = N$, the GNAR model becomes the classic first-order vector autoregression (VAR) type model with covariates, for which the number of parameters will quickly explode as $N$ increases.

Another popular strategy to model high dimensional time series is to impose some structural assumptions on the autoregression coefficient matrix of the VAR model. Examples include
assuming that the autoregression coefficient matrix is sparse  \citep{basu2015regularized,zhu2020nonconcave,nicholson2020high} or has a low rank structure \citep{negahban2011estimation,basu2019low,wang2022high}.
However, the aforementioned approaches do not incorporate the observed network structure for the model estimation and therefore can be less efficient when such information is available, which is demonstrated through simulation studies in Section~\ref{sec:compare}.
In addition, we remark that the model~\eqref{gnar} assumes that individuals are influenced in a similar way by friends of the same type they follow in a network. Although this assumption is reasonable for sparse networks, it is difficult to hold true in densely connected networks. In such situations, alternative high-dimensional VAR models~\cite[e.g.,][]{basu2019low} may be more suitable.


	Recently, modeling heterogeneity among individuals by imposing group structures has received considerable attention in panel data literature. For example, \cite{bonhomme2015grouped} considered grouped time-varying fixed effects for linear panel model and \cite{bester2016grouped}
	demonstrated that grouped individual fixed effects may improve the model estimation. 	\cite{ando2016panel} introduced grouped factor structure for linear panel data models. \cite{su2016identifying}
	proposed a Classifier Lasso (C-Lasso) procedure
	for simultaneous group identification and parameter estimation of panel data models. 
\cite{zhang2019quantile} studied clustering of panel data using quantile regression.
	More recently, \cite{liu2020identification} revisited the estimation and inference for the grouped panel data model with a possibly over-specified number of groups.  Similar structures are also used in~\cite{fang2020group}.   As we shall elaborate further, due to the existence of the network structure and the time-invariant covariates in model~\eqref{gnar}, the theoretical investigation of the GNAR model faces additional challenges compared to existing panel data models.

	\subsection{Comparison to existing works}	
	
A simplified version of model~\eqref{gnar} is considered in \cite{zhu2018grouped}, where they assume that $\beta_{g_ig_j} = \beta_{g_i}$ for any $g_j$'s, which is less realistic for network data. We wish to remark that our work is fundamentally different from \cite{zhu2018grouped}. Firstly, the model in \cite{zhu2018grouped} is essentially a finite Gaussian mixture model, for which group membership estimation consistency of network nodes cannot be established. In contrast, our work treats nodal group memberships as parameters that can be consistently estimated. Secondly, the asymptotic normality in \cite{zhu2018grouped} is established under the assumption that the true nodal memberships are known while our theory takes into account the potential group membership estimation errors. Thirdly, \cite{zhu2018grouped} assumes that the number of latent groups $G$ is known. In our work, not only do we allow $G$ to be over-specified but also give a data-driven method for consistently choosing $G$. Finally, our much stronger theoretical results are established without imposing restrictive assumptions on the network structure as those in~\cite{zhu2018grouped}; see Conditions~\ref{assum:prop} and~\ref{assum:prop1} for details. This further significantly expands the applicability of the proposed method.

{Our work is also significantly different from
the community network autoregression (CNAR) model recently proposed in \cite{chen2020community}, where they utilize the concept of ``community" that arises from
the community detection literature \citep{rohe2011spectral,lei2015consistency}. 	Although the ``group" structure in our work  appears to share some similarities  with ``community", they are fundamentally different.
The ``community" is typically determined by the connectivities among different network nodes, and the community structure is used to model the generating mechanism of the network structure, and the network structure is assumed to be random in~\cite{chen2020community}.  In contrast, for our GNAR model, the network structure is treated as deterministic over time, which is a reasonable framework for many applications and has been frequently used, see, e.g., \cite{fox2016modeling, farajtabar2017coevolve, zhu2017network, zhu2018network}.
Furthermore, unlike the ``community" in~\cite{chen2020community}, the groups of network nodes in our  GNAR model are primarily determined by node-specific characteristics, i.e., $\nu_{g_i}$'s and $\bzeta_{g_i}$'s.
In addition, we consider time invariant covariates $\bz_i$
instead of time dependent covariates in~\cite{chen2020community}.
Therefore, while modeling network time series data with similar structures, the research focus and theoretical challenges in our work is  fundamentally different from those in~\cite{chen2020community}.
}

Our theoretical findings appear to have a similar flavor as those in \cite{liu2020identification}. However, the technical proofs are significantly different, primarily due to the introduction of (1) the network effects $\beta_{g_i g_j}$'s, and (2) the time-invariant covariates $\bz_i$'s in model~\eqref{gnar}. Firstly, in~\cite{liu2020identification}, once the model parameters are estimated, the estimated membership $\wh g_i$ does not depend on values of other $\wh g_j$'s owning to the independence between different individuals in panel data. However, because of  $\beta_{g_i g_j}$'s in model~\eqref{gnar}, even when model parameters are given, the estimated $\wh g_i$ will inevitably depend on the estimated memberships of its connected nodes. The interplay between $\wh g_i$'s significantly complicated our theoretical investigations compared to those in \cite{liu2020identification}. Secondly, for panel data considered in~\cite{liu2020identification}, all model parameters related to an individual $i$ can be consistently estimated by using only the time series data from the $i$th individual given a sufficiently large $T$. However, this is not the case when we have time-invariant covariates $\bz_i$'s, in which case the fixed effects  $\bzeta_{g}$'s can only be consistently estimated by pooling data from all nodes in Group $g$. This is especially difficult since the true group memberships are unknown. To address these two challenges, we developed a new set of technical tools in the proof. As a result, although our Theorem~\ref{thm_member} only establishes convergence rates in probability, which is weaker than the almost sure convergence obtained in~\cite{liu2020identification}, it does provide more insights on how the network structure impacts the convergence rates. To establish asymptotic normality, we also proposed a refinement algorithm for the estimated group memberships that is not needed in~\cite{liu2020identification}.

	\subsection{Main Contributions and Organization}

	The main contributions of our work can be summarized as follows. First, we propose a highly interpretable GNAR model that is suitable for modeling multivariate time series observed on a network with heterogeneous nodes. Second, we give detailed conditions under which both model parameters and node memberships in the GNAR model can be consistently estimated, even if the number of groups $G$ is over-specified. Third, we propose an information criterion that can consistently choose the true number of groups when $N, T\to\infty$. Lastly, we show that, under suitable conditions, if the number of groups is correctly specified,
	the estimated model parameters converge to a multivariate normal distribution at a convergence rate of $\sqrt{NT}$, which enables valid statistical inference based on the proposed GNAR model.
	
	The rest of the paper is organized as follows. Section~\ref{sec:con} gives details on the proposed methodology including model description, computational algorithm, and sufficient conditions to establish estimation consistency when the number of latent groups $G$ is over-specified.  Section~\ref{sec:infer} establishes the asymptotic normality of the model parameter estimators when $G$ is correctly specified. Extensive simulation studies are conducted in Section~\ref{sec:numerical} and real data applications are given in Section~\ref{sec:data}. Details on the initialization of the proposed algorithm is given in the Appendix. All technical proofs and additional simulation studies are collected in the supplementary material.

	{\bf Notations.}
	Denote by $\bI_n$ the identity matrix with $n\times n$ dimension.
	Define $[G] = \{1,\cdots, G\}$
	and $[G]^n = \{ (g_1,\cdots, g_n)^\top: g_i\in [G]\}$.
	For an arbitrary vector $\bv = (v_1,\cdots, v_n)^\top\in \mR^n$, denote
	the $L_2$-norm  as $\|\bv\| = (\sum_{i = 1}^n v_i^2)^{1/2}$
	and $L_\infty$-norm as $\|\bv\|_\infty = \max_{1\le i\le n}|v_i|$.
For any set $\mS$, denote $|\mS|$ as its cardinality.
Finally, $\|\M\|_F = \tr\{\M^\top\M\}^{1/2}$ denotes the Frobenius norm of matrix $\M$.

	\section{Model Estimation}
	\label{sec:con}
	For a given number of groups $G$, denote the membership vector as $\mG = (g_1,\cdots, g_N)^\top\in[G]^N$.
	Define $\btheta = (\btheta_1^\top,\cdots, \btheta_G^\top)^\top\in\mR^{G(p+1)}$ with $\btheta_g = (\nu_g, \bzeta_g^\top)^\top\in \mR^{p+1}$, and $\bbeta = (\bbeta_1^\top,\cdots, \bbeta_G^\top)^\top\in \mR^{G^2}$ with $\bbeta_g=(\beta_{g1}, \beta_{g2},\cdots, \beta_{gG})^\top$ for $g\in[G]$.
	Correspondingly, the true parameters are defined as
	$\bnu^0 = (\nu_1^0,\cdots, \nu_{G_0}^{0\top})^\top\in \mR^{G_0}$,
	$\bzeta^0 = (\bzeta_{1}^0,\cdots, \bzeta_{G_0}^0)^\top \in \mR^{G_0\times p}$,
	and $\bbeta^0 = (\bbeta_1^{0\top},\cdots, \bbeta_{G_0}^{0\top})^\top\in \mR^{G_0^2}$ with $\bbeta_g^0=(\beta_{g1}^0, \beta_{g2}^0,\cdots, \beta_{gG_0}^0)^\top\in \mR^{G_0}$, where $G_0$ is the true number of groups.
	The membership vector $\mG$ as well as parameters $\btheta$ and $\bbeta$ can be estimated by minimizing the following quadratic loss function
	\begin{align}
	&Q(\btheta, \bbeta, \mG) =\frac{1}{N}\sum_{i = 1}^NQ_i(\btheta,\bbeta,\mG), \label{Q_obj1}
	\end{align}
	where $Q_i(\btheta, \bbeta, \mG) = T^{-1}\sum_{t = 1}^T\big(Y_{it}- \sum_{j = 1}^N\beta_{g_ig_j}w_{ij}Y_{j(t-1)} - \nu_{g_i} Y_{i(t-1)}  - \bz_i^\top\bzeta_{g_i}\big)^2$, for $i=1,\cdots,N.$ If $\mG$ is known, the optimization of $Q(\btheta, \bbeta, \mG)$ with respect to $\btheta$ and $\bbeta$ is convex and has a closed-form solution. However, we need to estimate $\mG$ jointly with other parameters, which makes the optimization of \eqref{Q_obj1} non-convex. In the next subsection, we give an iterative algorithm to minimize~\eqref{Q_obj1}.
	
	\subsection{An Optimization Algorithm}

	Note that the loss function~\eqref{Q_obj1} can be written as
	\begin{align}
	&Q(\btheta, \bbeta, \mG) =\sum_{g=1}^G\left\{\frac{1}{NT}\sum_{i = 1}^N\sum_{t = 1}^T\Big(Y_{it}- \cX_{i(t-1)}^\top\bxi_g \Big)^2 I(g_i=g)\right\}, \label{Q_obj1-new}
	\end{align}	
	where the vector $\cX_{i(t-1)} = (\wt Y_{i(t-1),1},\cdots, \wt Y_{i(t-1),G},Y_{i(t-1)},\bz_i^\top)^\top\in \mR^{G+p+1}$ with  $\wt Y_{i(t-1),g'} = \sum_{j = 1}^N w_{ij}Y_{j(t-1)}I(g_j = g')$ and $\bxi_g = (\bbeta_g^\top, \btheta_g^\top)^\top\in \mR^{G+p+1}$ for any $g,g'\in[G]$.
	It is straightforward to see that  $\bxi_g$'s can be estimated separately when $\mG$ is given.
	Specifically, let $\bX_g$ and $\bY_g$ be the design matrix and the response vector obtained by stacking all $\cX_{i(t-1)}^\top$'s and $Y_{it}$'s with $g_i=g$ and $1\le t\le T$, respectively. Then for a given $\mG$, the minimizer of $Q(\btheta, \bbeta, \mG)$ is of the following form
	\beq
	\wh \bxi_g = \big(\bX_g^\top\bX_g\big)^{-1}\big(\bX_g^\top\bY_g\big),\quad g=1,\cdots,G.\label{upsilon_g}
	\eeq
	
	Based on~\eqref{Q_obj1} and~\eqref{upsilon_g}, we propose the following iterative algorithm to minimize $Q(\btheta, \bbeta, \mG)$ with rerespect to $\btheta,\bbeta$ and $\mG$ jointly.
	
	\begin{itemize}
		\item [(a)] Obtain an initial membership estimator $\wh\mG^{(0)}$ using the  $k$-means algorithm given in the Appendix. Use $\wh\mG^{(0)}$ and \eqref{upsilon_g} to find initial estimators $\wh\btheta^{(0)}$ and $\wh\bbeta^{(0)}$.
		\item [(b)] {\bf Update group memberships:} in the $(k+1)$th iteration, update  each entry of $\wh\mG^{(k)}$
		sequentially, where $\wh \mG^{(k)}$ is the membership estimator in the $k$th step.
		Specifically, the group membership of node $i$ is updated by
		\begin{align}
		\wh g_i^{(k+1)} = \argmin_{g\in[G]} Q\Big(\wh\btheta^{(k)}, \wh\bbeta^{(k)}, \wh\mG_{-i}(g)\Big), \label{upG}
		\end{align}
		where $\wh\mG_{-i}(g)=\big(\wh g_1^{(k+1)}, \cdots,\wh g_{i-1}^{(k+1)}, g,\wh g_{i+1}^{(k)},\cdots,\wh g_{N}^{(k)}\big)^\top$, $i=1,\cdots,N.$ Repeat~\eqref{upG} for $i = 1,\cdots,N$ until no change can be made for $\wh\mG^{(k+1)}$.
		\item [(c)] {\bf Update the parameter estimates:}
		fix the group membership $\wh\mG^{(k+1)}$, and obtain the updated parameter estimates $\wh\btheta^{(k+1)}$ and $\wh\bbeta^{(k+1)}$  using (\ref{upsilon_g}).
		\item [(d)] Repeat (b)--(c) until the convergence criterion is met.
	\end{itemize}
	
The above optimization algorithm is a $k$-means type algorithm which consists of two major steps.
The first step is that we update the group memberships given the model parameters.
The second step is that we update the model parameters given the group memberships.
The algorithm framework is adopted by several group panel data models in recent literature \citep{ando2016panel,ando2017clustering,zhang2019quantile,liu2020identification}.
The main difference between our algorithm and the other approaches mainly lies in the first step due to introducing the network structure. Specifically,
	in the first step, when updating $\wh g_i$, we need to fix group memberships of all nodes that follow the node $i$ due to the existence of the network effect parameters $\beta_{g_ig_j}$'s in~\eqref{Q_obj1}.
On the contrary, in classical group panel data models, one can update the group membership $g_i$ separately for $i = 1,\cdots, N$ since the independence is typically assumed among the individuals.
For a given initial membership estimator $\wh\mG^{(0)}$, the above algorithm converges rather fast. However, since $Q(\btheta, \bbeta, \mG)$ is non-convex, it is important to search the solution with multiple initial values to escape from local minimums. In the Appendix, we propose an algorithm to search for multiple $\wh\mG^{(0)}$'s using a set of $k$-means algorithms, which works sufficiently well for all our numerical examples.
We prove that the algorithm can attain local convergence,
where the details are given in Appendix \ref{appen:local_conv} in the supplementary material.

	\subsection{Conditions for Estimation Consistency}	
	
	The GNAR model (\ref{gnar}) can be written in a vector form as following
	\beq
	\by_t = \bB \by_{t-1} +\bmu_z + \bve_t,\quad t=1,\cdots,T,\label{gnar_vec}
	\eeq
	where $\by_t = (Y_{1t},\cdots, Y_{Nt})^\top$, $\bmu_z = (\bz_1^\top \bzeta_{g_1}, \cdots,\bz_N^\top \bzeta_{g_N})^\top$, $\bve_t = (\ve_{1t},\cdots, \ve_{Nt})^\top$, and
	$\bB$ is an $N\times N$ matrix whose $(i,j)$th entry is $b_{ij} = w_{ij}\beta_{g_ig_j}$ for $i\ne j$ and $b_{ii} = \nu_{g_i}$ for $i,j=1,\cdots,N$. We next give sufficient conditions for estimation consistency.
	
	Suppose that the true number of latent groups is $G_0$ and the true group memberships are given by $\mG^0=(g_1^0,\cdots,g_N^0)^\top$ with $g_i^0\in[G_0]$.
For each node $i$, we denote $\mN_i=\{j: a_{ij}\ne 0\}$ as the set of the nodes that the node $i$ follows.

	\begin{assumption}\label{assum:eps_dist}
		({\sc Distribution})
		Assume that $\ve_{it}$, $1\le i\le N, 1\le t\le T$, are independent identically distributed (i.i.d.) zero-mean
		sub-Gaussian random variables with a scale factor $0<\sigma_1<\infty$, that is
		$E\{\exp(u\ve_{it})\} \le  \exp(\sigma_1^2u^2/2)$ for any $u$.
		Assume that $\bz_i$'s are fixed covariates
		satisfying $\max_{1\le i\le N}\|\bz_i\|_\infty<\infty$.
	\end{assumption}

	\begin{assumption}\label{assum:group_sep}
		{(\sc True Parameters)}
		Assume that (a) $\max\limits_{1\le g,g'\le G_0}\{|\beta_{gg'}^0| +  |\nu_g^0|\}<1$; (b) there exists a constant $c_0>0$ such that $\min\limits_{g\ne g'\in[G_0]}\big\{|\nu_{g}^0 - \nu_{g'}^0|^2 +
		\|\bzeta_{g}^0 - \bzeta_{g'}^0\|^2\big\}\ge c_0$.
	\end{assumption}
	
	\begin{assumption}\label{assum:prop}
		{\sc (Network Structure A)}
		For any $g,g'\in[G_0]$, define proportions $\pi_{g,N}=N^{-1}\sum_{i=1}^{N}I(g_i^0=g)$ and
		$ \pi_{gg',N}=N^{-1}\sum_{i=1}^N n_i^{-1}\sum_{j\in\mN_i} {I(g_i^0=g,g_j^0=g')}$.
		Assume that there exist $\pi_g$ and $\pi_{g g'}$ such that  $\pi_{g,N}\to\pi_g$ and
		$\pi_{g g',N}  \rightarrow \pi_{g g'}$ as $N\to\infty$, and that
		there exists a constant $c_{\pi}>0$ such that $\min\limits_{g, g'\in[G_0]}\min\{\pi_g, \pi_{gg'}\}\ge c_{\pi}$.
	\end{assumption}

	Condition \ref{assum:eps_dist} assumes that the innovations follow a  sub-Gaussian distribution,
	which is commonly used in high dimensional data analysis
	\citep{wang2013calibrating,lugosi2019sub,fan2021augmented}.
	Condition~\ref{assum:group_sep} (a) is a mild sufficient condition to ensure the stationarity of the vector autoregression model~\eqref{gnar_vec}, which is similar to  the stationarity condition of \cite{zhu2017network}. Condition~\ref{assum:group_sep} (b) requires that true parameters from different latent groups are sufficiently apart from each other, as similarly required by \cite{liu2020identification}. Condition~\ref{assum:prop} assumes that there are sufficiently number of nodes in each latent group, which is needed for consistent estimation of $\nu_g$'s and $\bzeta_g$'s. It also
{poses assumptions on the network structure, which basically}
requires that there are sufficient number of connected edges between any two groups to ensure consistent estimation of network effect parameters $\beta_{gg'}^0$ for $g,g'\in[G_0]$.
In addition, we provide local convergence result of the proposed numerical algorithm. The details are given in Appendix \ref{appen:local_conv}.

	\begin{assumption}\label{assum:para_space}
		{\sc (Parameter Space)}
		Assume that there exists a constant $R>0$ such that
		$\max\limits_{g\in[G]} \max\{|\nu_g|,\|\bbeta_g\|_{\infty},\|\bzeta_g\|_\infty\}\le R$.
	\end{assumption}
	
	\begin{assumption}\label{assum:sep_z}
		{(\sc Fixed-effect Identifiability)}
		Let $\mS_{g,N} = \{i: g_i^0 = g\}$ for $g\in [G_0]$.
			For any subset
			$ \mS_{g}'\subset \mS_{g,N}$ with $|\mS_{g}'|\ge c_0N^{\ve_z}$,
			it holds
			$|\mS_{g}'|^{-1}\lambda_{\min}(\sum_{i\in \mS_{g}'} \bz_i\bz_i^\top)\ge \tau_{\min}$ as $N\rightarrow \infty$,
			where $0<\ve_z  < 1$ and $\tau_{\min}>0$ are positive constants.

	\end{assumption}
	
	Condition~\ref{assum:para_space} assumes that the parameter space is compact, which is a standard condition in statistical theory. Condition~\ref{assum:sep_z} is a sufficient condition for the identifiability of fixed-effect parameters $\bzeta_g^0$, $g\in[G_0]$. It asserts that a sufficiently large set of nodes (i.e., greater than $c_0N^{\varepsilon_z}$) from any true group $g\in[G_0]$ should contain sufficient information to uniquely identify the corresponding fixed-effect vector $\bzeta_{g}$.  Note that Condition~\ref{assum:sep_z} trivially holds if there is only an intercept term in the fixed-effect, in which case $\bz_i\equiv 1$ for any $1\le i\le N$. In particular, when $\bz_i\equiv 1$, our theory still holds with $\ve_z=0$.

	 As we shall show in the next subsection, the convergence rate of model parameters is consequently affected by the value of $\varepsilon_z$.

	\subsection{Estimation Consistency with an Over-specified $G$}	
	We now establish the estimation consistency when $G\ge G_0$. Denote $(\wh\btheta,\wh\bbeta,\wh\mG)$ be the minimizer of~\eqref{Q_obj1} with $\wh\mG=(\wh g_1,\cdots,\wh g_N)^\top$.  To this end, we define the estimated groups as $\wh \mC_g = \{i: \wh g_i = g\}$ for $g\in[G]$ and a mapping $\chi: [G]\to[G_0]$ as
	\be
	\label{map0}
	\chi(g)=\argmax_{g'\in[G_0]}\sum_{i=1}^{N}I\left(i\in \wh \mC_g, g_i^0=g'\right),\quad g\in[G].
	\ee
	In other words, $\chi(g)$ gives the true membership of majority of nodes being assigned to $\wh\mC_g$ for any $g\in[G]$. The membership error rate can be consequently defined as
	\be
	\label{err}
	\wh \varrho_{NT}=\frac{1}{N}\sum_{g=1}^{G}\sum_{i=1}^{N}I\left(i\in\wh\mC_g, g_i^0\neq \chi(g)\right).
	\ee
	We remark that $1-\wh \varrho_{NT}$ gives the percentage of the nodes that are majority in all estimated groups, which is commonly referred to as the clustering purity \citep{schutze2008introduction}. 
	
Denote by $\overline{n}=\sqrt{N^{-1}\sum_{i=1}^N n_i^2}$ and $n_{\max}=\max_{1\le i\le N}n_i$ as the average and maximum of the out-degree of all network nodes. For a given $T$, we define the following quantity
	\be
	\label{nup}
 n_{up}=\inf_{C\ge 1}\Big\{C:\frac{1}{NC^2}\sum_{i=1}^N I(n_i>C)\le \frac{(\ol n+\log (N))^2}{T}\Big\}.
	\ee
 It readily follows that $1\leq n_{up}\le n_{\max}$.  For a sufficiently large $T$, \eqref{nup} implies that only a small fraction of nodes can follow more than $n_{up}$ network nodes. In this sense, $n_{up}$ serves as a measure of the network connectivity upper bound for a given $T$ to ensure estimation consistency,
	 and it is involved in the consistency result as stated in
	the following Theorem.

	\bet\label{thm_member}
	Assume Conditions \ref{assum:eps_dist}--\ref{assum:sep_z} and that $ n_{up}{\{\ol n+\log (N)\}/\sqrt{T}}\rightarrow 0$ as $(N,T)\rightarrow\infty$. Given a fixed $G\ge G_0$, it holds that\\
	(a). $\wh \varrho_{NT}=O_p\left( n_{up}^2\{\ol n+\log (N)\}^2/T\right)+O_p\left(N^{-1+\varepsilon_z}\right)$,\\
	(b). $N^{-1}\sum_{i=1}^N|\wh\nu_{\wh g_i} - \nu_{g_i^0}^0|^2+ N^{-1}\|\wh\bB - \bB^0\|_F^2  = O_p\left(n_{up}^2\{\ol n+\log (N)\}^2/T\right)$,\\
	(c). $N^{-1}\sum_{i=1}^N \big\|\wh\bzeta_{\wh g_i} - \bzeta_{g_i^0}^0\big\|^2 = O_p\left(n_{up}^2\{\ol n+\log (N)\}^2/T+N^{-1+\varepsilon_z}\right)$,\\
	where $\bB^0$ and $\wh\bB$ are the true and estimated autoregression matrices as defined in~\eqref{gnar_vec}.
	\eet
	The proof is given in the supplementary material.
	
Theorem \ref{thm_member} (a) asserts that the fraction of network nodes that are assigned to an incorrect group approaches $0$ as $N,T\to\infty$, In particular, ignoring the $ O_p\left(N^{-1+\varepsilon_z}\right)$ term, the rate of convergence in part (a) is mainly controlled by $T$ rather than $N$.
This is consistent with our observations in the simulation study, where an increase in $T$ results in a large reduction in $\wh \varrho_{NT}$ while a larger $N$ only yields a marginal decrease or even an increase of $\wh \varrho_{NT}$.
	The convergence rates given in Theorem~\ref{thm_member}~(b)--(c) are of the same form, suggesting that to compensate for the impacts of network effects as well as the network dependence structure, one needs a larger $T$ by a factor of $n_{up}^2$ (assuming $n_{\max}<\log(N)$)  to ensure estimation consistency compared to the case when all nodes are isolated without any followers.
Consequently, the  result is different from existing results from the panel data literature when the individuals are typically treated as independent such as \cite{liu2020identification}.
Particularly, the network structure related quantities (i.e., $n_{up}, \ol n$) are not incorporated.
Moreover, compared to the network data setting considered by~\cite{zhu2018grouped}, we remark that  while our theoretical results are more sophisticated, our theory imposes much fewer restrictions on the network structure, see Conditions~\ref{assum:prop} and~\ref{assum:prop1} for details.
	
	\subsection{Consistent Selection of $G_0$}
	Although the consistency results in Theorem~\ref{thm_member} can apply to any $G\ge G_0$, it is still of practical interest to identify the true value of $G_0$ since a smaller $G$ can improve the model interpretability and estimation accuracy. In particular, as we will show in Section~\ref{sec:infer}, valid statistical inference can be performed if $G_0$ is consistently identified.
	This motivates us to design a data-driven selection criterion for $G$.
	
	With a slight abuse of notations, denote $\wh \btheta^{(G)},\wh\bbeta^{(G)},\wh\mG^{(G)}$ as the estimated model parameters and group memberships when the number of groups is specified as $G$. The optimal $\wh G$ is chosen by minimizing the following group information criterion (GIC)
	\begin{align}
	\mbox{GIC}_{\lambda_{NT}}(G) = \log \big\{Q\big(\wh \btheta^{(G)},\wh\bbeta^{(G)},\wh\mG^{(G)}\big)\big\} + \lambda_{NT}G,\label{qic}
	\end{align}
	where $\lambda_{NT}>0$ is a tuning parameter.
	In the following theorem, we show that if $\lambda_{NT}$ is appropriately chosen,
	the GIC can identify the true number of groups $G_0$ consistently.

	\bet\label{thm_selectionG}
	Assume Conditions \ref{assum:eps_dist}--\ref{assum:sep_z} and that  $n_{up}{\{\ol n+\log (N)\}/\sqrt{T}}\rightarrow 0$ as $(N,T)\to\infty$. If $\lambda_{NT}$ satisfies following conditions
	\beq
	\lambda_{NT}n_{up}\rightarrow 0 \text{ and }	\lambda_{NT}^{-1} \left(n_{up}\{\ol n+\log (N)\}^2/T\right)\rightarrow 0,
	\eeq
	then we have that $P(\wh G = G_0)\rightarrow 1$ as $(N,T)\to\infty$.
	\eet
	The proof is given in the supplementary material.
	
The GIC is designed in the similar fashion of the BIC in the model selection literature \citep{chen2008extended,zou2009adaptive,wang2013calibrating}. Some discussion on the condition $\lambda_{NT}n_{up}\rightarrow 0$ is in order. In our proof of Theorem~\ref{thm_selectionG}, we manage to show that if $G<G_0$, one has that $\Delta_{NT}=Q\big(\wh \btheta^{(G)},\wh\bbeta^{(G)},\wh\mG^{(G)}\big)-Q\big(\wh \btheta^{(G_0)},\wh\bbeta^{(G_0)},\wh\mG^{(G_0)}\big)>c/n_{up}$ for some constant $c>0$. In panel data models, it is typically true that $\Delta_{NT}>c$ for some constant $c>0$ if $G<G_0$, see, e.g., \cite{liu2020identification}. The difference is due to the existence of the network effects $\beta_{g_ig_j}$'s in~\eqref{Q_obj1}, in which case the bias caused by the smaller parameter space (due to a smaller $G$) is offset by the extra flexibility arising from the network effects, leading to the extra $n_{up}$ term in $\Delta_{NT}$. As a result, we require $\lambda_{NT}n_{up}\rightarrow 0 $ in contrast to $\lambda_{NT}\rightarrow 0$ suggested in, e.g., \cite{liu2020identification}.

	
	
	\section{Model Inference}
	\label{sec:infer}
	We next investigate the asymptotic distribution of the model parameter estimators. Compared to Section~\ref{sec:con}, we need to further assume $G=G_0$ as in~\cite{liu2020identification} and the following additional identifiability condition to Condition~\ref{assum:group_sep}.
	\begin{assumption}\label{assum:group_sep1}
		{(\sc Group Identifiability)}
		There exists a positive constant $c_0$ such that
		$\min_{g\ne g'\in[G_0]} \Big\{|\nu_{g}^0 - \nu_{g'}^0| +
		\min_{1\le i\le N}|\bz_i^\top(\bzeta_{g}^0 - \bzeta_{g'}^0)|\Big\}\ge c_0$.
	\end{assumption}	
		\begin{assumption}\label{assum:prop1}
		{\sc (Network Structure B)}
		For any $g,g'\in[G_0]$, there exist a constant $c_0>0$ such that
		$N^{-1}\sum_{i=1}^N {n_i^{-2}}\sum_{j\in\mN_i} {I(g_i^0=g,g_j^0=g')}\geq c_0$.
		\end{assumption}

	Condition~\ref{assum:group_sep1} requires that two latent groups either have different momentum effect parameters, i.e., $\nu_g$'s, or different fixed-effect parameters, i.e., $\bzeta_g$'s, that can separate any two nodes in the network.
Recall that we require that $\bz_i$ always includes the intercept term. Specifically, if $p = 1$ (i.e., $\bz_i = 1$ for $1\le i\le N$), Condition \ref{assum:group_sep1} reduces to
$\min_{g\ne g'\in[G_0]} \{|\nu_{g}^0 - \nu_{g'}^0| +
		|\bzeta_{g}^0 - \bzeta_{g'}^0|\}\ge c_0$. In more general cases, it is slightly more restrictive than the Condition~\ref{assum:group_sep} but still reasonable for many applications.
Condition~\ref{assum:prop1} is a slightly more restrictive condition on the network structure than Condition~\ref{assum:prop}, which is the price to pay to achieve the asymptotic normality of parameter estimators. It implies that the number of nodes with bounded out-degrees should be of the order $O(N)$, suggesting that the network density should not be too high.
Our Lemma~\ref{lem:Sig_g_lowerbound} in the supplement also shows that Condition~\ref{assum:prop1} ensures all diagonal elements of the matrix $\bSigma^{(g)}$ in~Theorem~\ref{thm_normal} to be greater than a constant $c>0$, which is necessary for $\bSigma^{(g)}$ to be strictly positive definite as assumed.
Compared to the network structure conditions of \cite{zhu2018grouped}, both Conditions~\ref{assum:prop} and~\ref{assum:prop1} are much simpler and more transparent.
	
	\subsection{Membership Refinement}
	To establish the asymptotic normality, we further propose an algorithm to refine the estimated group memberships. 	Denote  by $\mG_i = (g_j: j\in \mN_i)^\top$ the group memberships of the nodes that the node $i$ follows and
	$\bvarphi_{g_i,\mG_i} = (n_i^{-1/2}\beta_{g_ig_j}: j \in \mN_i)^\top$, for $i=1,\cdots,N$. Then the loss function corresponding to the node $i$, i.e., $Q_i(\btheta, \bbeta, \mG)$ in~\eqref{Q_obj1}, can also be written as a function of  $\btheta_{g_i}$ and $\bvarphi_{g_i, \mG_i}$, denoted by
	$Q_i(\btheta_{g_i}, \bvarphi_{g_i, \mG_i})$. Note that $Q_i(\btheta_{g_i}, \bvarphi_{g_i, \mG_i})$ does not only depend on its own membership $g_i$ but also memberships of its neighbors $\mG_i$. As a result, the minimizer of the loss function~\eqref{Q_obj1}, denoted as  $(\wh\btheta,\wh\bbeta,\wh\mG)$, does not necessarily minimize each $Q_i(\btheta_{g_i}, \bvarphi_{g_i, \mG_i})$, which creates a hurdle for
	analyzing the asymptotic distribution of  $(\wh\btheta,\wh\bbeta,\wh\mG)$. To circumvent this difficulty, we propose a refinement of the estimated memberships $\wh\mG$ using an approximate node-specific profile loss function.	 Specifically, let $ \wh\bPhi_i  =\{ (n_i^{-1/2}\wh\beta_{g_ig_j}: j \in \mN_i)^\top: g_i \in[G],\mG_i= (g_j: j\in \mN_i)^\top\in [G]^{n_i}\}$
	with $\wh\beta_{g_ig_j}$'s being the corresponding entries in $\wh\bbeta$ obtained from minimizing~\eqref{Q_obj1}. Given  $\wh\bbeta$ , $\wh\bPhi_i$ is the collection of all possible estimated network effects between the node $i$ and the nodes it follows (i.e., $\mN_i$), obtained by exhausting membership assignments to nodes $i$ and nodes in $\mN_i$. The approximate node-specific profile loss function of $g_i$ is defined as
	\[
	Q_i^P(g)=\min_{\bvarphi_i\in \wh\bPhi_i} Q_i(\wh\btheta_{g}, \bvarphi_i), \quad g\in[G], i=1,\cdots,N.
	\]
	 The definition of $Q_i^P(g)$ eliminates the impacts of membership estimates for nodes in $\mN_i$ when determining $g_i$, which facilitates our technical proofs.
	Define the optimal $\wh g_i^\dag = \arg\min_{g\in [G]}Q_i^P(g)$, and if $Q_i^P({\wh g_i^\dag})$ is much smaller than
$Q_i(\wh\btheta_{\wh g_i}, \wh\bvarphi_{\wh g_i,\wh\mG_i})$, then we have reason to switch from the original estimated membership $\wh g_i$ to $\wh g_i^\dag$.  Consequently, we define the  refined estimated  membership  as following
	\beq
	\wh g_i^r=
	\begin{cases}
		\wh g_i, & \mbox{if }  Q_i(\wh\btheta_{\wh g_i}, \wh\bvarphi_{\wh g_i,\wh\mG_i}) -
Q_i^P({\wh g_i^\dag})\le \frac{1}{\sqrt{T}}Q_i^P({\wh g_i^\dag})\\
		\wh g_i^\dag, & \mbox{if }
		{Q_i(\wh\btheta_{\wh g_i}, \wh\bvarphi_{\wh g_i,\wh\mG_i})} - Q_i^P({\wh g_i^\dag}) > {\frac{1}{\sqrt{T}}Q_i^P({\wh g_i^\dag})}.\label{rfm}
	\end{cases}
	\eeq

Intuitively,  \eqref{rfm} asserts that one should only switch the membership from $\wh g_i$ to $\wh g_i^\dag$ if the reduction of the loss at the node $i$ is more than $T^{-1/2}\times 100\%$ of the minimum possible profile loss. We shall show in the next subsection that such a refinement strategy ensures the asymptotic normality of the resulting parameter estimators.		
	\begin{remark}
Our simulation study in Section~4.1 shows that the refined estimator performs slightly worse than the unrefined estimator in most case scenarios, although the differences are rather small. Given this observation, we wish to remark that the membership refinement algorithm serves as more of a device that facilitates our theoretical investigations and can be skipped in the practical use of the proposed method.
	\end{remark}

	
	
	\subsection{Asymptotic Normality}
	In this section, we establish asymptotic normality for model parameter estimators when $G=G_0$. The first challenge is to obtain a stronger convergence result for the membership mis-classification rate than Theorem~\ref{thm_member} (a). Denote by $\wh\mG^r=(\wh g_1^r,\cdots,\wh g_N^r)^\top$  the refined estimated memberships using~\eqref{rfm}, and $\wh\mC_g^r=\{i: \wh g_i^r=g\}$, $g\in[G_0]$ as the estimated clusters. The following Theorem gives the uniform consistency of the parameter estimators as well as the group membership estimators.
	\bet\label{thm_consistency}
	Assume Conditions \ref{assum:eps_dist}--\ref{assum:prop1} and that
	$n_{\max}^2\xggrev{n_{up}}{\{n_{\max}+\log (N)\}/\sqrt{T}}\rightarrow 0$.	Then if $G=G_0$, as $(N,T)\rightarrow\infty$, it holds that,\\
	(a). $\sup_{1\le i\le N} \Big\{|\wh\nu_{\wh g_i^r} - \nu_{g_i^0}^0|^2 + |\bz_i^\top\wh\bzeta_{\wh g_i^r} - \bz_i^\top \bzeta_{g_i^0}^0|^2\Big\} = o_p\left( 1/({n_{\max}n_{up}})\right)$;\\
	(b). for $1\le g\le G$, there exists one $1\le g'\le G_0$, such that
	$P(\wh \mC_g^r= \mC_{g'}^0)\rightarrow 1$.
	\eet
	The proof is given in the supplementary material.

	Theorem \ref{thm_consistency} can be viewed as an enhanced version of Theorem~\ref{thm_member} for the special case with $G=G_0$, which states that under some regularity conditions, all group memberships can be correctly estimated (subject to a label permutation) with a probability tending to $1$ as $(N, T)\to\infty$. Similar results have also been established in the panel data literature \citep[e.g.,][]{liu2020identification}. However, similar to Theorem~\ref{thm_member}, special care must be paid to the network structure in our work by considering the network structure related factors as
$n_{up}$ and $n_{\max}$.

	Making use of 	Theorem \ref{thm_consistency} (b), the following Theorem establishes the asymptotic normality of model parameter estimators when $G=G_0$.
	\begin{theorem}\label{thm_normal}
		Let $\wh \bxi_g^r$ be defined by (\ref{upsilon_g}) with the refined membership
		$\wh \mG^r$ and $\bxi_g^0$ be the corresponding true parameter vector (after an appropriate label permutation).
Define $\bSigma^{(g)}=\lim_{(N_g,T)\to\infty}(N_gT)^{-1}E\big(\bX_g^{0\top}\bX_g^0\big)$ with $\bX_g^0$ as in (\ref{upsilon_g}) by plugging in the true membership $\mG^0$. 	Assume that $G = G_0$, Conditions in Theorem \ref{thm_consistency} hold, and that $\bSigma^{(g)}$ is strictly positive definite.  Then, it holds that
		\begin{align*}
		\sqrt{N_gT} \big(\wh \bxi_g^r - \bxi_g^0\big)\xrightarrow{d}
		N\big(0, \sigma^2(\bSigma^{(g)})^{-1}\big),\quad g\in[G_0], \label{eq:thm_normal}
		\end{align*}
		where $N_g=\sum_{i=1}^N I(g_i^0=g)$.
	\end{theorem}
	The proof is given in the supplementary material.
	
	Theorem \ref{thm_normal} states that for each $g\in[G_0]$, $\wh \bxi_g^r$ is $\sqrt{N_gT}$ consistent for $\bxi_g^0$
	with an asymptotic covariance matrix given by $\sigma^2(\bSigma^{(g)})^{-1}$.
	The asymptotic covariance is the same as the oracle estimator \eqref{upsilon_g} which knows the group membership in advance.
In practice, we can estimate $\bSigma^{(g)}$ using the refined memberships $\{\wh g_i^r: i\in [N]\}$. Specifically, we use $\wh \bSigma^{(g)} = (\wh N_g^rT)^{-1} \bX_g^{r\top}\bX_g^r$, where  $\bX_g^r = (\cX_{i(t-1)}: \wh g_i^r = g, t\in [T])^\top$ and $\wh N_g^r = \sum_{i = 1}^N I(\wh g_i^r = g)$.
	In addition, we estimate $\sigma^2$ by
	$\wh \sigma^2 = (NT)^{-1} \sum_{i,t}(Y_{it} - \cX_{i(t-1)}^\top \wh \bxi_{\wh g_i^r}^r)^2$.
	By Condition \ref{assum:prop}, we have that $N_g=N\pi_{g,N}\ge c_\pi N$,
	which suggests that $N_g$ diverges in the same order of $N$.
Theorem~\ref{thm_normal} enables us to conduct valid statistical inference for model parameters, including the momentum effects ($\nu_g^0$'s), the network effects ($\beta_{gg'}^0$'s),  and the fixed-effects ($\bzeta_g^0$'s).
	which is supported by the numerical results given in  Section~\ref{sec:numerical}.

	\section{Simulation Studies}
	\label{sec:numerical}
	
	To demonstrate the finite sample performance of the proposed method, we conduct a number
	of simulation studies with different network structures
	and parameter settings using model~\eqref{gnar}.
	For all settings, the time-invariant covariate  $\bz_i = (z_{i1},\cdots, z_{ip})^\top$'s are independently generated from
	a multivariate normal distribution $N(0, \bI_p)$ with $p  =2$.
	The innovation term $\ve_{it}$'s
	are independently sampled from $N(0,1)$.
	For each network structure, we consider two settings with $G_0 = 2$ and $G_0 = 3$, and  sample the memberships of the network nodes from multinomial  distribution
		with a $(\pi_1, \pi_2) = (0.5, 0.5)$ for $G_0 = 2$
		and $(\pi_1, \pi_2, \pi_3) = (0.3, 0.3, 0.4)$ for $G_0 = 3$ respectively.
	We consider two network structures.

	{\sc 1. Stochastic Block Model (SBM)}.
	For this network structure, the nodes are partitioned into $C$ communities. If nodes $i$ and $j$ belong to the same community, the chance of them being connected is set as $P(a_{ij} = 1) = 2\log(N)/N$, otherwise the chance reduces to $P(a_{ij} = 1) = \log(N)/N$.  This corresponds to the challenging case where the exact recovery of the communities are not possible \citep{abbe2020entrywise}. For different network sizes $N = 100, 200, 300$, we set $C =5, 10, 20$
	respectively.
	
	{\sc 2. Power-Law Distribution Network}.
	In this network, the node in-degrees ($d_i=\sum_{j=1}^n a_{ji}$) follow a power-law distribution, which is suitable for social networks where the majority of nodes have few followers but a small percent of nodes have a large number of followers.
	Following \cite{clauset2009power}, we generate the network structure as follows.
	First, for each node $i$, we generate $\wt d_i$
	by $P(\wt d_i = k) \propto k^{-2.5}$ and set the in-degree of the node as $d_i = 4 \wt d_i$.
	Next, for the $i$th node, we randomly pick $d_i$ nodes as its followers.

	\begin{table}[ht!]
		\caption{True parameters for $G_0 = 2, 3$.}\label{tab_G}
		\begin{center}
			\scalebox{0.7}{
				\begin{tabular}{c|ccc|c|cc|ccc|c|cc}
					\hline
					&\multicolumn{6}{c|}{$G_0 = 2$} & \multicolumn{6}{c}{$G_0 = 3$}\\
					\hline
					&                \multicolumn{3}{c|}{$\beta_{gg'}^0$} & $\nu_g^0$ & \multicolumn{2}{c|}{$\bzeta_g^0$} &
					\multicolumn{3}{c|}{$\beta_{gg'}^0$} & $\nu_g^0$ & \multicolumn{2}{c}{$\bzeta_g^0$}\\
					$g/g'$&                1&2&3& - & -&-&
					1&2&3 & - & -&-\\
					\hline
					1&0.3 & -0.2 & - & 0.4 & -0.8 & 0.8 & 0.15 & 0.2 & -0.1 & 0.2 & -1.2 & 0.4 \\
					2&0.1 & 0.3 & - & 0.6& -0.32 &1.2 & 0.1 & 0.3 & -0.2 & 0.4 & -0.8 & 0.8 \\
					3&- &- &- & - & - &- & 0.15 & 0.1 & 0.3 & 0.6 & -0.32 & 1.2 \\
					\hline
				\end{tabular}
			}
		\end{center}
	\end{table}

	For each of the two network structures, the performances of the proposed method are evaluated under three parameter settings.
	In {\sc Scenario 1}, we specify the true parameters as in Table \ref{tab_G} respectively for $G_0 = 2$ and $G_0 = 3$.
	In {\sc Scenario 2}, we set $\nu_1^0 =\cdots =\nu_{G_0}^0 = {0.4}$, in which case the groups only differ in network effect parameters $\beta_{gg'}^0$'s and fixed-effect parameters $\bzeta_g^0$'s.
	Lastly, in {\sc Scenario 3}, we set $\bzeta_1^0=\cdots=\bzeta_{G_0}^0 = 0$ and groups only differ in network effect parameters $\beta_{gg'}^0$'s  and momentum parameters $\nu_g^0$'s.


	\subsection{Estimation and Inference when $G=G_0$}
	\label{sec:G0}
	
	When $G=G_0$, we consider both of the unrefined and refined estimators. 	Denote the estimates obtained from the proposed algorithm
		as $\wh \bbeta^{(b)}$, $\wh \bnu^{(b)}$, and $\wh \bzeta^{(b)}$ for the $b$th simulation run {and let $\wh \bbeta^{r(b)}$, $\wh \bnu^{r(b)}$, and $\wh \bzeta^{r(b)}$ be the corresponding estimates after the refinement. The group membership estimation error rate is computed as
		$\wh \varrho_{NT} = B^{-1}\sum_{b=1}^B \wh \varrho_{NT}^{(b)}$, where $\wh \varrho_{NT}^{(b)}$ is obtained by applying definition~\eqref{err} to the $b$th simulation run.} The estimation accuracy can be directly measured by the root mean squared error (RMSE) of $\wh \bbeta$, $\wh \bnu$, and $\wh \bzeta$ after a suitable label permutation. For example, for $\bbeta^0$, the RMSE is defined as RMSE$_{\bbeta}=B^{-1}\sum_{b=1}^B \|\wh \bbeta^{(b)} - \bbeta^0\|$.
To evaluate the performance of statistical inference using Theorem~\ref{thm_normal}, we construct $95\%$ confidence interval for each model parameter {based on the refined estimates}.
Taking $\nu^0$ as an example, in the $b$th simulation run, we construct
	$95\%$ confidence interval for $\nu_g^0$ as
	CI$_{g}^{(b)}= (\wh \nu_g^{r(b)} - 1.96\wh \se_g^{(b)}, \wh \nu_g^{r(b)} + 1.96\wh \se_g^{(b)})$,
	where $\wh\se_g^{(b)}$ is the estimated asymptotic standard error based on Theorem~\ref{thm_normal}.
	The average error in coverage probability (AE$_{\rm cp}$) for all components in $\bnu^0$ is then calculated as
 AE$_{\rm cp,\bnu} = G_0^{-1}\sum_{g=1}^{G_0}|B^{-1}\sum_{b=1}^B I(\nu_g^0\in \mbox{CI}_g^{(b)}) - 0.95|$. 
The AE$_{\rm cp}$'s for $\bbeta$ and $\bzeta$ are similarly defined.  Finally, for a direct comparison, we compute the same measures for the oracle estimators obtained when the true group memberships are known, denoted as $\wh \bbeta_o$, $\wh \bnu_o$, and $\wh \bzeta_o$.

	\begin{table}[htp]
		\caption{RMSE's ($\times 10^{-2}$) and AE$_{\rm cp}$'s (\%, in the parenthesis)  in {\sc Scenario 1} for the SBM network.}\label{tab_sbm_type0}
		\begin{center}
			\scalebox{0.55}{%
				\begin{tabular}{c|cc|ccc|cccc|cccc}
					\hline
					\hline
&	&      & \multicolumn{3}{c|}{Oracle Estimator} & \multicolumn{4}{c|}{GNAR without Refinement}& \multicolumn{4}{c}{GNAR with Refinement} \\
$G_0$	&$N$ &$T$ & $\wh\bbeta_o$ & $\wh\bnu_o$ & $\wh\bzeta_o$& $\wh\bbeta$  & $\wh\bnu$ & $\wh\bzeta$&$\wh \varrho_{NT} (\%)$& $\wh\bbeta^r$  & $\wh\bnu^r$ & $\wh\bzeta^r$ & $\wh \varrho_{NT} (\%)$ \\
\hline
& 100  &  100  &  3.94 (0.45) & 1.61 (1.50) & 4.93 (1.85)  &  4.20 (1.20) & 1.67 (2.30) & 5.10 (2.45)  &  2.90  &  4.20 (1.20) & 1.67 (2.30) & 5.10 (2.45)  &  2.90 \\
&  &  200  &  2.76 (0.45) & 1.14 (1.10) & 3.45 (1.20)  &  2.84 (0.85) & 1.17 (1.50) & 3.52 (1.70)  &  0.86  &  2.84 (0.85) & 1.17 (1.50) & 3.52 (1.70)  &  0.86 \\
&  &  300  &  2.24 (1.30) & 0.90 (1.10) & 2.77 (0.60)  &  2.26 (1.35) & 0.92 (0.90) & 2.81 (0.85)  &  0.31  &  2.26 (1.35) & 0.92 (0.90) & 2.81 (0.85)  &  0.31 \\
2 & 200  &  100  &  2.33 (0.60) & 1.06 (1.20) & 3.30 (0.65)  &  2.48 (1.40) & 1.09 (0.70) & 3.35 (0.80)  &  2.28  &  2.47 (1.40) & 1.09 (0.70) & 3.35 (0.80)  &  2.28 \\
&  &  200  &  1.63 (0.75) & 0.72 (2.20) & 2.29 (1.00)  &  1.66 (0.65) & 0.75 (1.50) & 2.32 (0.35)  &  0.62  &  1.66 (0.65) & 0.75 (1.50) & 2.32 (0.35)  &  0.62 \\
&  &  300  &  1.32 (1.00) & 0.62 (0.50) & 1.91 (0.70)  &  1.32 (0.75) & 0.63 (0.40) & 1.93 (0.80)  &  0.22  &  1.32 (0.75) & 0.63 (0.40) & 1.93 (0.80)  &  0.22 \\
& 300  &  100  &  2.01 (0.60) & 0.85 (1.10) & 2.69 (1.60)  &  2.10 (1.95) & 0.90 (0.90) & 2.78 (1.30)  &  2.51  &  2.10 (1.95) & 0.90 (0.90) & 2.78 (1.30)  &  2.51 \\
&  &  200  &  1.38 (0.85) & 0.60 (1.30) & 1.89 (0.70)  &  1.40 (0.85) & 0.61 (1.10) & 1.90 (0.65)  &  0.74  &  1.40 (0.85) & 0.61 (1.10) & 1.90 (0.65)  &  0.75 \\
&  &  300  &  1.12 (0.45) & 0.48 (0.80) & 1.52 (0.80)  &  1.13 (0.50) & 0.49 (0.50) & 1.52 (0.90)  &  0.26  &  1.13 (0.50) & 0.49 (0.50) & 1.52 (0.90)  &  0.26 \\
\hline
& 100  &  100  &  12.87 (0.60) & 2.77 (0.53) & 7.38 (0.80)  &  15.07 (3.00) & 2.98 (1.73) & 8.07 (1.57)  &  3.08  &  15.50 (3.98) & 3.04 (2.33) & 8.21 (1.97)  &  3.28 \\
&  &  200  &  8.99 (0.96) & 1.93 (1.27) & 5.18 (0.77)  &  9.35 (1.02) & 1.98 (1.47) & 5.34 (1.00)  &  0.78  &  9.41 (1.11) & 1.98 (1.40) & 5.38 (1.00)  &  0.82 \\
&  &  300  &  7.52 (0.69) & 1.52 (0.53) & 4.18 (0.90)  &  7.61 (0.69) & 1.55 (0.87) & 4.23 (0.97)  &  0.27  &  7.63 (0.71) & 1.55 (0.87) & 4.24 (0.97)  &  0.28 \\
3 & 200  &  100  &  6.93 (0.64) & 1.84 (0.53) & 4.92 (0.83)  &  7.83 (3.09) & 1.95 (1.27) & 5.16 (1.37)  &  2.92  &  8.17 (4.33) & 2.03 (3.20) & 5.32 (1.97)  &  3.19 \\
&  &  200  &  4.73 (0.40) & 1.27 (0.73) & 3.45 (0.50)  &  4.92 (0.73) & 1.30 (0.47) & 3.52 (0.87)  &  0.77  &  4.96 (0.93) & 1.31 (0.87) & 3.53 (1.03)  &  0.83 \\
&  &  300  &  3.89 (0.53) & 1.08 (0.47) & 2.90 (0.87)  &  3.95 (0.40) & 1.09 (1.00) & 2.92 (0.87)  &  0.30  &  3.96 (0.38) & 1.09 (1.00) & 2.92 (0.90)  &  0.30 \\
& 300  &  100  &  5.37 (1.11) & 1.42 (0.40) & 3.87 (1.10)  &  5.93 (1.89) & 1.51 (0.40) & 4.10 (1.03)  &  2.90  &  6.16 (3.51) & 1.69 (3.73) & 4.37 (2.67)  &  3.21 \\
&  &  200  &  3.85 (0.60) & 1.02 (0.93) & 2.77 (0.60)  &  3.94 (0.93) & 1.05 (0.67) & 2.84 (0.77)  &  0.78  &  3.98 (1.40) & 1.07 (1.00) & 2.87 (0.93)  &  0.82 \\
&  &  300  &  3.13 (0.80) & 0.83 (0.53) & 2.25 (0.63)  &  3.16 (0.84) & 0.85 (0.33) & 2.27 (0.77)  &  0.27  &  3.17 (0.87) & 0.85 (0.33) & 2.28 (0.67)  &  0.28 \\
					\hline
					\hline
				\end{tabular}
			}
		\end{center}
		\caption{RMSE's ($\times 10^{-2}$) and AE$_{\rm cp}$'s (\%, in the parenthesis)  in {\sc Scenario 2} for the SBM network.}\label{tab_sbm_type1}
			\begin{center}
		\scalebox{0.54}{%
			\begin{tabular}{c|cc|ccc|cccc|cccc}
				\hline
				\hline
				&	&      & \multicolumn{3}{c|}{Oracle Estimator} & \multicolumn{4}{c|}{GNAR without Refinement}& \multicolumn{4}{c}{GNAR with Refinement} \\
				$G_0$	&$N$ &$T$ & $\wh\bbeta_o$ & $\wh\bnu_o$ & $\wh\bzeta_o$& $\wh\bbeta$  & $\wh\bnu$ & $\wh\bzeta$&$\wh \varrho_{NT} (\%)$& $\wh\bbeta^r$  & $\wh\bnu^r$ & $\wh\bzeta^r$ & $\wh \varrho_{NT} (\%)$ \\
				\hline
\hline
& 100  &  100  &  4.27 (0.90) & 1.62 (0.40) & 4.44 (0.50)  &  6.09 (10.10) & 2.20 (11.00) & 5.90 (8.55)  &  10.32  &  6.09 (10.10) & 2.20 (11.00) & 5.90 (8.55)  &  10.32 \\
&  &  200  &  3.07 (0.60) & 1.20 (0.30) & 3.18 (0.55)  &  3.88 (5.80) & 1.40 (4.30) & 3.61 (3.65)  &  6.35  &  3.88 (5.80) & 1.40 (4.30) & 3.61 (3.65)  &  6.35 \\
&  &  300  &  2.55 (0.50) & 0.96 (0.40) & 2.57 (1.00)  &  3.05 (5.45) & 1.12 (4.30) & 3.41 (3.55)  &  5.26  &  3.05 (5.45) & 1.12 (4.30) & 3.41 (3.55)  &  5.26 \\
2 & 200  &  100  &  2.74 (0.85) & 1.17 (0.30) & 3.08 (0.25)  &  3.91 (12.15) & 1.59 (10.00) & 4.04 (7.25)  &  10.55  &  3.90 (12.10) & 1.59 (10.10) & 4.04 (7.30)  &  10.55 \\
&  &  200  &  1.91 (0.50) & 0.82 (0.40) & 2.17 (0.55)  &  2.30 (6.65) & 0.91 (3.10) & 2.41 (2.50)  &  6.41  &  2.30 (6.65) & 0.91 (3.10) & 2.41 (2.50)  &  6.41 \\
&  &  300  &  1.58 (0.80) & 0.67 (0.50) & 1.77 (0.45)  &  1.74 (2.75) & 0.73 (3.40) & 1.89 (2.45)  &  4.62  &  1.74 (2.75) & 0.73 (3.40) & 1.89 (2.45)  &  4.62 \\
& 300  &  100  &  2.23 (1.00) & 0.93 (0.80) & 2.49 (1.00)  &  3.37 (12.25) & 1.30 (10.50) & 3.41 (7.20)  &  9.72  &  3.37 (12.25) & 1.30 (10.50) & 3.41 (7.25)  &  9.72 \\
&  &  200  &  1.60 (0.60) & 0.65 (0.70) & 1.72 (0.70)  &  2.12 (7.15) & 0.82 (4.60) & 2.38 (3.15)  &  6.50  &  2.12 (7.10) & 0.82 (4.60) & 2.38 (3.15)  &  6.50 \\
&  &  300  &  1.32 (0.50) & 0.53 (0.50) & 1.40 (0.90)  &  1.70 (6.85) & 0.67 (3.70) & 2.21 (3.35)  &  5.26  &  1.70 (6.85) & 0.67 (3.70) & 2.21 (3.35)  &  5.26 \\
\hline
& 100  &  100  &  12.21 (0.84) & 2.62 (0.53) & 7.40 (0.70)  &  29.41 (22.38) & 3.91 (10.87) & 17.49 (15.43)  &  15.16  &  29.22 (22.53) & 3.95 (10.93) & 17.47 (15.77)  &  15.21 \\
&  &  200  &  8.59 (0.51) & 1.93 (0.73) & 5.32 (0.70)  &  17.06 (14.31) & 2.43 (5.40) & 10.08 (8.87)  &  8.22  &  16.95 (14.13) & 2.45 (5.73) & 10.06 (8.83)  &  8.21 \\
&  &  300  &  7.07 (1.00) & 1.58 (0.93) & 4.44 (1.73)  &  11.85 (9.62) & 1.81 (3.67) & 7.00 (6.73)  &  5.26  &  11.82 (9.51) & 1.83 (3.80) & 6.93 (6.57)  &  5.22 \\
3 & 200  &  100  &  6.53 (0.71) & 1.85 (0.53) & 4.88 (0.43)  &  10.98 (15.33) & 2.59 (11.13) & 7.19 (9.27)  &  11.51  &  11.16 (16.29) & 2.61 (11.47) & 7.23 (9.83)  &  11.63 \\
&  &  200  &  4.65 (0.71) & 1.31 (1.40) & 3.46 (0.70)  &  6.16 (7.18) & 1.45 (2.73) & 4.03 (3.63)  &  5.26  &  6.23 (7.44) & 1.45 (3.20) & 4.04 (3.57)  &  5.29 \\
&  &  300  &  3.82 (0.82) & 1.06 (0.47) & 2.77 (0.77)  &  4.70 (5.80) & 1.12 (1.80) & 3.19 (2.30)  &  3.35  &  4.71 (5.78) & 1.13 (1.73) & 3.19 (2.33)  &  3.34 \\
& 300  &  100  &  5.30 (0.87) & 1.45 (0.93) & 3.97 (0.73)  &  8.18 (13.62) & 2.01 (9.53) & 5.67 (8.37)  &  10.39  &  8.28 (13.93) & 2.03 (9.47) & 5.74 (8.63)  &  10.53 \\
&  &  200  &  3.72 (0.56) & 1.01 (0.53) & 2.77 (0.83)  &  4.90 (7.96) & 1.19 (3.53) & 3.84 (4.37)  &  5.72  &  4.94 (8.22) & 1.20 (3.67) & 3.87 (4.80)  &  5.74 \\
&  &  300  &  3.05 (0.67) & 0.82 (0.87) & 2.26 (0.80)  &  3.66 (4.49) & 0.93 (2.67) & 2.93 (2.97)  &  3.71  &  3.70 (4.87) & 0.94 (2.87) & 2.93 (2.87)  &  3.70 \\
				\hline
				\hline
			\end{tabular}
		}
	\end{center}
	\end{table}

	\begin{table}[htp]
		\caption{RMSE's ($\times 10^{-2}$) and AE$_{\rm cp}$'s (\%, in the parenthesis)  in {\sc Scenario 3} for the SBM network.}\label{tab_sbm_type2}
		\begin{center}
	\scalebox{0.54}{%
		\begin{tabular}{c|cc|ccc|cccc|cccc}
			\hline
			\hline
			&	&      & \multicolumn{3}{c|}{Oracle Estimator} & \multicolumn{4}{c|}{GNAR without Refinement}& \multicolumn{4}{c}{GNAR with Refinement} \\
			$G_0$	&$N$ &$T$ & $\wh\bbeta_o$ & $\wh\bnu_o$ & $\wh\bzeta_o$& $\wh\bbeta$  & $\wh\bnu$ & $\wh\bzeta$&$\wh \varrho_{NT} (\%)$& $\wh\bbeta^r$  & $\wh\bnu^r$ & $\wh\bzeta^r$ & $\wh \varrho_{NT} (\%)$ \\
\hline
& 100  &  100  &  7.59 (0.60) & 1.63 (0.60) & 2.81 (0.75)  &  11.47 (13.80) & 2.28 (13.30) & 3.47 (5.80)  &  13.45  &  11.47 (13.80) & 2.28 (13.30) & 3.47 (5.80)  &  13.45 \\
&  &  200  &  5.51 (1.40) & 1.10 (0.60) & 2.00 (0.75)  &  6.34 (4.20) & 1.31 (5.60) & 2.15 (1.90)  &  5.23  &  6.34 (4.20) & 1.31 (5.60) & 2.15 (1.90)  &  5.23 \\
&  &  300  &  4.52 (0.55) & 0.92 (0.30) & 1.60 (0.80)  &  4.83 (1.10) & 0.99 (2.10) & 1.65 (1.25)  &  2.21  &  4.83 (1.10) & 0.99 (2.10) & 1.65 (1.25)  &  2.21 \\
2 & 200  &  100  &  5.18 (0.50) & 1.08 (1.20) & 1.96 (0.40)  &  7.86 (14.20) & 1.89 (20.90) & 2.43 (6.25)  &  12.80  &  7.86 (14.20) & 1.89 (20.90) & 2.43 (6.25)  &  12.80 \\
&  &  200  &  3.65 (1.15) & 0.75 (0.50) & 1.36 (0.70)  &  4.24 (4.00) & 0.92 (5.30) & 1.45 (2.15)  &  5.12  &  4.24 (4.00) & 0.92 (5.30) & 1.45 (2.15)  &  5.12 \\
&  &  300  &  2.97 (0.95) & 0.61 (0.90) & 1.13 (0.90)  &  3.16 (1.30) & 0.65 (2.20) & 1.16 (1.35)  &  2.18  &  3.16 (1.30) & 0.65 (2.20) & 1.16 (1.35)  &  2.18 \\
& 300  &  100  &  4.30 (0.65) & 0.87 (0.80) & 1.56 (0.80)  &  6.60 (15.75) & 1.72 (25.50) & 1.95 (6.15)  &  12.78  &  6.60 (15.75) & 1.72 (25.50) & 1.95 (6.15)  &  12.78 \\
&  &  200  &  3.05 (1.10) & 0.63 (0.90) & 1.09 (0.65)  &  3.51 (4.75) & 0.78 (6.20) & 1.18 (1.60)  &  4.97  &  3.51 (4.75) & 0.78 (6.20) & 1.18 (1.60)  &  4.97 \\
&  &  300  &  2.46 (0.55) & 0.50 (0.30) & 0.90 (0.70)  &  2.57 (1.35) & 0.55 (1.60) & 0.93 (0.70)  &  2.08  &  2.57 (1.35) & 0.55 (1.60) & 0.93 (0.70)  &  2.08 \\
\hline
& 100  &  100  &  21.24 (0.67) & 2.76 (0.27) & 4.76 (0.90)  &  42.53 (23.29) & 6.62 (26.60) & 6.50 (9.17)  &  18.64  &  42.18 (23.04) & 6.62 (26.73) & 6.50 (9.20)  &  18.70 \\
&  &  200  &  15.08 (0.89) & 1.88 (0.27) & 3.32 (0.40)  &  18.83 (4.42) & 2.42 (6.93) & 3.66 (2.13)  &  4.56  &  18.85 (4.44) & 2.42 (6.87) & 3.66 (2.17)  &  4.58 \\
&  &  300  &  12.41 (0.67) & 1.55 (0.53) & 2.74 (0.80)  &  13.49 (1.71) & 1.74 (2.80) & 2.84 (1.17)  &  1.76  &  13.50 (1.73) & 1.74 (2.80) & 2.84 (1.17)  &  1.77 \\
3 & 200  &  100  &  13.69 (0.73) & 1.82 (0.73) & 3.03 (0.80)  &  23.95 (18.44) & 3.33 (20.00) & 4.09 (8.70)  &  14.71  &  23.75 (17.93) & 3.32 (19.93) & 4.10 (8.70)  &  14.78 \\
&  &  200  &  9.60 (0.76) & 1.27 (0.20) & 2.14 (0.77)  &  11.39 (3.80) & 1.58 (6.07) & 2.32 (1.83)  &  4.73  &  11.37 (3.84) & 1.59 (6.07) & 2.33 (1.87)  &  4.75 \\
&  &  300  &  7.81 (0.69) & 1.06 (1.13) & 1.79 (1.00)  &  8.38 (1.31) & 1.15 (2.33) & 1.84 (1.50)  &  1.89  &  8.38 (1.31) & 1.15 (2.33) & 1.84 (1.53)  &  1.89 \\
& 300  &  100  &  10.71 (0.42) & 1.49 (0.87) & 2.46 (0.83)  &  18.33 (17.89) & 2.62 (18.47) & 3.31 (8.63)  &  13.93  &  18.13 (17.29) & 2.61 (18.73) & 3.32 (8.83)  &  13.99 \\
&  &  200  &  7.57 (0.60) & 1.09 (1.80) & 1.77 (0.53)  &  8.92 (3.71) & 1.34 (6.73) & 1.94 (2.17)  &  4.55  &  8.91 (3.69) & 1.34 (6.80) & 1.94 (2.17)  &  4.56 \\
&  &  300  &  6.15 (1.13) & 0.87 (0.53) & 1.44 (0.70)  &  6.52 (1.67) & 0.96 (3.13) & 1.49 (1.10)  &  1.87  &  6.52 (1.69) & 0.96 (3.20) & 1.49 (1.13)  &  1.87 \\
			\hline
			\hline
		\end{tabular}
	}
\end{center}
	\end{table}
	Summary statistics based on $B=500$ simulation runs are given in Tables \ref{tab_sbm_type0}--\ref{tab_sbm_type2} for the SBM network.   Simulation results for the power-law network yield rather similar conclusions and are given in the supplementary material. First,
	Tables \ref{tab_sbm_type0}--\ref{tab_sbm_type2} suggest that as either $N$ or $T$ increases,
	the parameter estimation accuracy consistently improves and approaches the estimation accuracy of the oracle estimators. However, the group membership estimation error rate $\wh\varrho_{NT}$ only gains a significant reduction when $T$ increases, which is consistent with our theoretical findings in Theorem~\ref{thm_member} (a). As $N, T$ increases,  the overall performance of the proposed method is much better with a $G_0=2$ than $G_0=3$, which is as expected. Second, we can see that  in {\sc Scenarios 2-3}, the $\wh\varrho_{NT}$'s are much higher compared to those of {\sc Scenario 1} because group separations are much greater in {\sc Scenario 1}. The important message from {\sc Scenarios 2-3} is that even if two groups only differ in either momentum parameters or fixed-effect parameters, they can be consistently separated using the proposed method given large enough $N$ and $T$.
	
	Finally, the differences between the unrefined and refined estimators appear to be rather small. In particular, when $G=2$, no membership switch was executed based on Algorithm~\eqref{rfm} since the clustering is relatively easier in this case. When $G=3$, the refined memberships appear to have slightly higher clustering errors in most case scenarios with only a few exceptions. Nevertheless, in all case scenarios, the  AE$_{\rm cp}$ values
		are rather small for both unrefined and refined estimators, suggesting that all confidence intervals have right nominal coverage probability when $N$ and $T$ are large. We can observe that the performances of the proposed estimators gradually approach those of the oracle estimators as $N$ and $T$ increase. This leads to further supports  our theoretical findings in Theorem~\ref{thm_normal}.
	
	\subsection{Estimation and Group Selection when $G\ge G_0$}
	\label{sec:compare}
	In this section, we evaluate the performance of the proposed method when the number of groups is mis-specified. The true number of groups is fixed at $G_0=3$.
Under this case, to measure the estimation accuracy, we use  the following criteria. For the estimation of $\bzeta^0$ and $\bnu^0$, we define
	RMSE$_{\bzeta,all} = (NB)^{-1}
	\sum_{i=1}^N\sum_{b=1}^B\|\wh \bzeta_{\wh g_i}^{(b)} - \bzeta_{g_i^0}^0\|$
	and
	RMSE$_{\bnu,all} =  (NB)^{-1}
	\sum_{i=1}^N\sum_{b=1}^B|\wh \nu_{\wh g_i}^{(b)} - \nu_{g_i^0}^0|$.
	For $\bbeta^0$, we define
	RMSE$_{\bbeta, all} = (NB)^{-1}
	\sum_{i=1}^N\sum_{b=1}^B\|\wh \bB_{i\cdot}^{(b)} - \bB_{i\cdot}^0\|$, which evaluates the estimation accuracy of the autoregression matrix
	$\bB^0$ defined in model~\eqref{gnar_vec}. Furthermore, we also evaluate the selection accuracy for number of groups using the GIC criterion proposed in~\eqref{qic}, for which we set the tuning parameter as
	$\lambda_{NT} = N^{1/10}T^{-1/2}/(2\min\{10, n_{0.9}\})$,
	where $n_{0.9}$ is the 90\% quantile of nodal out-degrees $\{n_i:1\le i\le N\}$.
	We compute the model selection rate (MSR) as
	MSR$(G) = B^{-1}\sum_{b=1}^B I (\wh G^{(b)} = G)$, for any given $G$,
	where $\wh G^{(b)}$ denotes the selected number of groups with the GIC in the $b$th
	simulation run.
{Specifically, MSR(3) corresponds to the percentage that the GIC correctly identifies the true group number $G_0 = 3$.}

For comparisons, we investigate the performances of several existing methods on the data generated from our model~\eqref{gnar}, including
the sparse VAR model by \cite{basu2015regularized}, and the grouped network autoregression model by \cite{zhu2018grouped}. For the sparse VAR model, we use the \textsf{fitVAR} function in the R package \textsf{sparsevar}.
However, since the sparse VAR model in \cite{basu2015regularized} does not include  time-invariant covariates as the model~\eqref{gnar}, we apply the method proposed in \cite{basu2015regularized} to centered time series $Y_{it}-T^{-1}\sum_{t=1}^{T}Y_{it}$ ($i=1,\cdots,N$) to eliminate the impacts of $\bz_i^\top \bzeta_{g_i}$'s, and focus on the estimation of $\bbeta^0$ and $\bnu^0$. For the grouped network autoregression model proposed in \cite{zhu2018grouped}, we implement both the EM algorithm (EM) and the two-step estimation method (TS), for which we set $G=G_0$. Summary statistics based on $B= 500$ simulation runs are given in Table~\ref{tab_overG}.

	\begin{table}[ht!]
		\caption{Simulation results for the SBM network  with varying $G$'s.}\label{tab_overG}
		\begin{center}
			\scalebox{0.6}{
				\begin{tabular}{ccc|ccccc|ccccc|ccccc}
					\hline
					\hline
					&     & & \multicolumn{5}{c|}{{\sc Scenario 1}} & \multicolumn{5}{c|}{{\sc Scenario 2}} & \multicolumn{5}{c}{{\sc Scenario 3}}  \\
					$N$ & $T$ & Method&
					$\wh\bbeta$ & $\wh\bnu$ & $\wh\bzeta$ & MSR & $\wh\varrho_{NT}$ &
					$\wh\bbeta$ & $\wh\bnu$ & $\wh\bzeta$ & MSR & $\wh\varrho_{NT}$ &
					$\wh\bbeta$ & $\wh\bnu$ & $\wh\bzeta$ & MSR & $\wh\varrho_{NT}$ \\
					
					&     & & \multicolumn{3}{c}{(RMSE$_{all}\times 10^{-2}$) } &\multicolumn{2}{c|}{(\%) }& \multicolumn{3}{c}{(RMSE$_{all}\times 10^{-2}$) } &\multicolumn{2}{c|}{(\%) }& \multicolumn{3}{c}{(RMSE$_{all}\times 10^{-2}$) } &\multicolumn{2}{c}{(\%) }  \\
					\hline
100  &  200  & Oracle  &  0.84 & 0.85 & 2.57  & - & - &  0.88 & 0.90 & 2.50  & - & - &  1.58 & 0.84 & 1.54  & - & -\\
& & GNAR-2 &  3.72 & 6.32 & 21.29  &  0.0  &  26.0 &  2.60 & 5.35 & 1.25  &  50.8  &  23.4 &  4.70 & 4.00 & 22.90  &  0.0  &  28.5\\
& & GNAR-3 &  0.92 & 0.95 & 2.87  &  100.0  &  0.5 &  2.28 & 1.76 & 1.66  &  49.2  &  4.3 &  1.58 & 0.99 & 5.44  &  96.8  &  4.3\\
& & GNAR-4 &  1.87 & 1.97 & 6.20  &  0.0  &  0.7 &  4.18 & 2.74 & 2.50  &  0.0  &  4.6 &  2.93 & 1.62 & 9.10  &  3.2  &  6.5\\
& & GNAR-5 &  2.70 & 2.68 & 8.54  &  0.0  &  1.2 &  5.33 & 3.55 & 2.97  &  0.0  &  6.1 &  4.13 & 2.17 & 13.63  &  0.0  &  9.6\\
&  &GNAR-$\hat G$ &  0.92 & 0.95 & 2.87 & - & -  &  2.44 & 3.58 & 1.45 & - & -  &  1.62 & 1.01 & 5.56 & - & - \\
&  & EM(G=3)   &  11.19 & 3.21 & 12.12  & -  &  6.7 &  10.19 & 3.14 & 2.48  & -  &  7.9 &  9.45 & 1.60 & 20.85  & -  &  20.7\\
&  & TS (G=3)  &  9.70 & 14.19 & 37.63  & -  &  22.5 &  10.52 & 6.04 & 2.48  & -  &  22.3 &  8.56 & 20.04 & 59.73  & -  &  40.2\\
&  & SparseVAR(1)  &  12.03 & 14.93 & - & - & -  &  11.96 & 14.92 & - & - & -  &  12.12 & 15.15 & - & - & - \\
\hline
100  &  300  & Oracle  &  0.68 & 0.67 & 2.07  & - & - &  0.71 & 0.73 & 2.04  & - & - &  1.30 & 0.69 & 1.27  & - & -\\
& & GNAR-2 &  3.68 & 6.26 & 21.23  &  0.0  &  26.0 &  2.25 & 5.15 & 1.03  &  15.6  &  22.5 &  4.58 & 4.01 & 22.38  &  0.0  &  28.0\\
& & GNAR-3 &  0.71 & 0.71 & 2.17  &  100.0  &  0.2 &  1.54 & 1.09 & 1.31  &  84.4  &  1.9 &  1.04 & 0.78 & 3.58  &  98.2  &  2.4\\
& & GNAR-4 &  1.47 & 1.61 & 5.09  &  0.0  &  0.2 &  3.10 & 1.90 & 1.98  &  0.0  &  1.9 &  2.00 & 1.33 & 6.18  &  1.8  &  3.7\\
& & GNAR-5 &  2.07 & 2.23 & 7.12  &  0.0  &  0.6 &  4.02 & 2.55 & 2.36  &  0.0  &  2.8 &  2.95 & 1.77 & 9.61  &  0.0  &  5.9\\
&  &GNAR-$\hat G$ &  0.71 & 0.71 & 2.17 & - & -  &  1.65 & 1.72 & 1.27 & - & -  &  1.06 & 0.79 & 3.63 & - & - \\
&  & EM(G=3)   &  11.16 & 2.96 & 11.37  & -  &  5.6 &  10.18 & 2.00 & 1.98  & -  &  3.6 &  9.38 & 1.39 & 20.08  & -  &  20.0\\
&  & TS (G=3)  &  9.93 & 12.38 & 32.34  & -  &  20.3 &  10.24 & 5.43 & 1.78  & -  &  20.2 &  8.54 & 18.94 & 56.23  & -  &  37.9\\
&  & SparseVAR(1)  &  10.45 & 11.66 & - & - & -  &  10.47 & 11.64 & - & - & -  &  10.60 & 11.75 & - & - & - \\
\hline
200  &  200  & Oracle  &  0.55 & 0.60 & 1.81  & - & - &  0.57 & 0.65 & 1.81  & - & - &  1.16 & 0.62 & 1.13  & - & -\\
& & GNAR-2 &  2.76 & 6.29 & 19.61  &  0.0  &  29.0 &  2.87 & 6.20 & 0.97  &  10.0  &  29.1 &  4.60 & 4.12 & 24.04  &  0.0  &  33.3\\
& & GNAR-3 &  0.64 & 0.72 & 2.14  &  100.0  &  0.5 &  1.85 & 1.56 & 1.20  &  90.0  &  4.4 &  1.59 & 0.72 & 6.10  &  99.0  &  5.8\\
& & GNAR-4 &  1.29 & 1.74 & 5.02  &  0.0  &  0.7 &  3.40 & 2.50 & 1.82  &  0.0  &  5.1 &  2.77 & 1.22 & 9.43  &  1.0  &  8.9\\
& & GNAR-5 &  1.91 & 2.66 & 7.33  &  0.0  &  1.3 &  4.63 & 3.61 & 2.31  &  0.0  &  7.0 &  3.79 & 1.99 & 13.12  &  0.0  &  12.0\\
&  &GNAR-$\hat G$ &  0.64 & 0.72 & 2.14 & - & -  &  1.95 & 2.02 & 1.18 & - & -  &  1.60 & 0.72 & 6.13 & - & - \\
&  & EM(G=3)   &  10.00 & 3.87 & 13.58  & -  &  9.5 &  9.26 & 2.91 & 1.82  & -  &  8.6 &  8.23 & 1.32 & 21.85  & -  &  25.2\\
&  & TS (G=3)  &  9.18 & 14.76 & 40.06  & -  &  27.1 &  9.70 & 7.19 & 1.53  & -  &  27.4 &  8.05 & 17.89 & 56.46  & -  &  49.1\\
&  & SparseVAR(1)  &  13.10 & 16.69 & - & - & -  &  13.12 & 16.70 & - & - & -  &  13.17 & 16.98 & - & - & - \\
\hline
200  &  300  & Oracle  &  0.45 & 0.51 & 1.51  & - & - &  0.47 & 0.52 & 1.43  & - & - &  0.94 & 0.51 & 0.95  & - & -\\
& & GNAR-2 &  2.69 & 6.19 & 19.35  &  0.0  &  28.8 &  2.49 & 5.97 & 0.79  &  0.0  &  28.3 &  4.46 & 4.19 & 23.61  &  0.0  &  32.8\\
& & GNAR-3 &  0.47 & 0.55 & 1.62  &  100.0  &  0.2 &  1.18 & 0.90 & 0.96  &  100.0  &  1.9 &  1.08 & 0.57 & 4.17  &  99.4  &  3.8\\
& & GNAR-4 &  0.98 & 1.44 & 4.13  &  0.0  &  0.3 &  2.43 & 1.66 & 1.44  &  0.0  &  2.1 &  1.93 & 0.93 & 6.55  &  0.6  &  5.8\\
& & GNAR-5 &  1.47 & 2.23 & 5.85  &  0.0  &  0.5 &  3.40 & 2.60 & 1.82  &  0.0  &  3.3 &  2.77 & 1.49 & 9.39  &  0.0  &  8.1\\
&  &GNAR-$\hat G$ &  0.47 & 0.55 & 1.62 & - & -  &  1.18 & 0.90 & 0.96 & - & -  &  1.09 & 0.57 & 4.18 & - & - \\
&  & EM(G=3)   &  10.00 & 3.72 & 13.15  & -  &  8.7 &  9.25 & 1.73 & 1.42  & -  &  3.9 &  8.22 & 1.21 & 21.49  & -  &  24.6\\
&  & TS (G=3)  &  9.39 & 12.94 & 35.35  & -  &  24.9 &  9.46 & 6.47 & 1.23  & -  &  24.7 &  8.09 & 17.32 & 54.57  & -  &  47.1\\
&  & SparseVAR(1)  &  11.19 & 13.20 & - & - & -  &  11.19 & 13.19 & - & - & -  &  11.26 & 13.19 & - & - & - \\
\hline
					\hline
				\end{tabular}
			}
		\end{center}
	\end{table}

	We first focus on the performance of the GNAR estimator. From Table~\ref{tab_overG}, we can observe that when the model is under-fitted ($G = 2$),
	the RMSE is much larger than when it is over-fitted ($G>3$) and the RMSE does not significantly decrease when both $N, T$ increase.
	This is in line with the fact that the under-fitted model leads to a significant model estimation bias.
	When $G$ is over-specified (i.e., $G>3$), we observe that the both RMSE and clustering error rate $\wh\varrho_{NT}$ are larger than
	those from the model with a correctly specified $G = 3$.
	That is due to the inflated model estimation uncertainty when the number of model parameters increases.  It may also be caused by the greater misclassification error with a larger $G$.
	In the meantime, the RMSE and $\wh\varrho_{NT}$ still decrease with an over-specified $G$
	as the sample size ($N$ and $T$) increases, which corroborates with the theoretical results in Theorem \ref{thm_member}. Finally, the MSR values are all close to $100\%$ for {\sc Scenario 1} and 3. For {\sc Scenario 2}, although it requires a much larger sample size $N$ and $T$, the MSRs also reach $100\%$ when $N,T$ are sufficiently large. This observation supports the selection consistency results given by Theorem~\ref{thm_selectionG}. In fact, in all case scenarios, when $G$ is  chosen by GIC as $\wh G$, the resulting RMSE$_{all}$'s are very close to those with a fixed $G=G_0=3$.

Among the competing methods, the SparseVAR(1) appears to have the worst estimation accuracies in terms of RMSE$_{\bnu, all}$ and RMSE$_{\bbeta, all}$, which is not surprising considering that the number of parameters to be estimated is of the order $O(N^2)$. Even with regularization, the estimation uncertainty can still be rather high. Between the EM and TS methods, it appears that the EM method consistently outperforms the TS method in terms of both estimation accuracy and clustering error $\wh\varrho_{NT}$. Finally, comparing the EM method to the proposed GNAR algorithm, we can see that the clustering errors are consistently higher for the EM method, especially in {\sc Scenario 1} and  {\sc Scenario 3}. Consequently, the estimation accuracies of the EM method also appear to be significantly worse than the GNAR estimator with either $G=3$ or $G=\wh G$ chosen by the GIC.  This observation suggests that if the network effects $\beta_{g_i g_j}$'s in model~\eqref{gnar} are misspecified as those in \cite{zhu2018grouped}, both model estimation and membership clustering will be negatively impacted.

\subsection{Performance under Misspecified Models}

In this section, we investigate the robustness of the GNAR model by studying its performance when the model is misspecified. For comparisons, we generate the data from the low-rank and structured vector auto-regressive  model \citep[LS-VAR,][]{basu2019low}
$
 	\y_t = \B\y_{t-1} + \bve_t, \text{ where } \B=\B_1+\B_2$,
 where $\B_1$ is a low-rank matrix and $\B_2$ is a sparse matrix. Compared to the GNAR model~\eqref{gnar_vec}, this model does not include any time-invariant covariates but employs an autoregressive coefficient matrix $\B$ of a specific structure.

In the following we consider two specifications for the $\B$ matrix.
In {\sc Case I}, we consider a sparse structure of $\B$.
In {\sc Case II}, we consider a low-rank+sparse structure of $\B$.
First, we consider a purely sparse case with $\B_1=\bm 0$ in {\sc Case I}. The sparse matrix is generated as $\B=\B_2=C_0(1-\rho)\A_1/\|\A_1\|_F+\rho \A_2/\|\A_2\|_F$, where $\A_1$ is a sparse matrix with around $5\%$ nonzero entries generated from a standard normal distribution and $\A_2$ is the coefficient matrix generated from the GNAR model with $G_0=3$ under {\sc Scenario 3} in Section~\ref{sec:G0}. The constant $C_0$ is chosen such that the spectral norm of $\B$ is $0.7$, and the ratio $\rho=0,0.3,0.5,0.7,1$.  We then apply the GNAR method and regularized estimation method proposed by \cite{basu2019low}  for LS-VAR to estimate the coefficient matrix $\B$ with various $N$ and $T$. The number of groups in the GNAR model is selected using the GIC \eqref{qic}. For the method proposed by \cite{basu2019low}, we use the {\sf fista.LpS} function in R package {\sf LSVAR}. The tuning parameters used by the {\sf fista.LpS} function are selected by minimizing the prediction error on a testing dataset with $T_{test} = 50$ using a model fitted by the remaining time points. After the tuning parameters are chosen, we refit the LS-VAR model with the whole data set.
 Summary statistics based on $B=500$ simulations are presented in Table~\ref{tab:compare2}, where we compute the relative estimation error (REE) as $\textup{REE} = B^{-1}
 \sum_{b = 1}^B \|\wh \B^{(b)} - \B^0\|_F/\|\B^0\|_F$ with $\wh \B^{(b)}$ being the estimated transition matrix in $b$th simulation run.

 In our simulation settings, the GNAR model is only correctly specified when $\rho=1$ while the LS-VAR model is always correct. Table~\ref{tab:compare2} shows that when $\rho\le 0.5$, the LS-VAR model performs much better than the GNAR model, suggesting that when the model misspecification is severe, the GNAR model produces large biases that make it much less accurate than more general models such as the LS-VAR model. However, when the model misspecification is not severe (e.g., $\rho=0.7$ or more), the GNAR may still outperform the LS-VAR model due to the benefit of the exploration of homogeneity. Such an observation suggests that the proposed GNAR model has some degree of robustness against model misspecification in the purely sparse case.

\begin{table}[htp]
	\caption{The REEs ($\times 10^2$) of the GNAR model and the LS-VAR (LS) model.}\label{tab:compare2}
	\begin{center}
					\scalebox{0.75}{
		\begin{tabular}{cc|cc|cc|cc|cc|cc}
			\hline
			\hline
			&  & \multicolumn{2}{c|}{ $ \rho =  0 $}  & \multicolumn{2}{c|}{ $ \rho =  0.3 $}  & \multicolumn{2}{c|}{ $ \rho =  0.5 $}  & \multicolumn{2}{c|}{ $ \rho =  0.7 $}  & \multicolumn{2}{c}{ $ \rho =  1 $} \\
			$N$ & $T$ & LS & GNAR & LS & GNAR & LS &GNAR & LS &GNAR & LS & GNAR\\
			\hline
					\multicolumn{12}{c}{Case I: Purely Sparse Structure} \\
			50  &  200 &  24.7  &  114.1 &  36.2  &  103.7 &  48.0  &  74.3 &  51.4  &  45.8 &  43.4  &  25.4\\
			50  &  400 &  17.1  &  115.6 &  25.2  &  104.8 &  33.8  &  73.5 &  37.0  &  45.1 &  33.4  &  23.7\\
			100  &  200 &  37.7  &  100.8 &  55.8  &  94.3 &  61.1  &  75.0 &  56.1  &  46.2 &  49.8  &  24.3\\
			100  &  400 &  27.0  &  100.6 &  41.5  &  93.8 &  46.0  &  74.2 &  42.4  &  44.3 &  34.9  &  21.1\\
\hline			
			\multicolumn{12}{c}{Case II: Low-rank+Sparse Structure} \\
			50  &  200 &  79.9  &  106.5 &  82.6  &  92.6 &  86.3  &  73.0 &  68.5  &  46.4 &  45.7  &  23.3\\
			50  &  400 &  61.5  &  101.9 &  68.0  &  88.3 &  68.8  &  68.8 &  51.8  &  42.9 &  33.4  &  18.2\\
			100  &  200 &  94.6  &  112.1 &  94.8  &  99.8 &  96.3  &  79.4 &  80.3  &  50.2 &  46.7  &  22.8\\
			100  &  400 &  79.4  &  106.3 &  82.7  &  94.2 &  84.9  &  74.1 &  63.2  &  46.4 &  33.7  &  19.0\\
				\hline
			\hline
		\end{tabular}
	}
	\end{center}
\end{table}

Next, in {\sc Case II}, we investigate the setting when $\B$ is not purely sparse (i.e., $\B_1\ne \bm 0$). In this case, we define $\B=C_0(1-\rho)\A_1^*/\|\A_1^*\|_F+\rho \A_2/\|\A_2\|_F$, where $\A_1^*$ is a symmetric $N\times N$ matrix with ${\rm Rank}(\A_1^*)=3$ and nonzero singular values as $1,1,1$, and $\A_2$ is generated the same way as in the purely sparse case.  The constant $C_0$ is chosen such that the spectral norm of $\B$ is $0.7$, and the ratio $\rho=0,0.3,0.5,0.7,1$.
In these settings, the GNAR model estimators ignore the low-rank part $C_2\A_1^*/\|\A_1^*\|_F$
and therefore are always biased except for the case $\rho=1$. We can see from Table~\ref{tab:compare2} that, the REEs of the GNAR model are rather similar to the purely sparse case. On the contrary, the REEs of the LS-VAR model deteriorate. One possible explanation is that to correctly recover the low-rank structure, the required $T$ should be much larger than those in the purely sparse case for a given $N$. Therefore, the estimation variance of the LS-VAR estimators becomes the dominant source of the estimation error, even exceeding the estimation bias of the GNAR estimators. It is reasonable to anticipate that for a given $N$, the performance of the LS-VAR estimator will improve as $T$ increases but the performance of the GNAR will stay roughly the same for $\rho<1$. Nevertheless, for finite $N$ and $T$, the GNAR model may still have good performance as long as the model is not severely misspecified.

\section{Real Data Examples}
\label{sec:data}

\subsection{Financial Contagion Analysis of Stock Market}
\label{sec:stock}
In this section, we study a data set that collects information on companies listed in the Chinese A share market in 2020.
It is common that many listed companies share a set of same shareholders and hence stock prices of these companies may correlate with each other. The shareholder network captures important inter-corporate dependence and has been an important research topic in financial risk management.   Companies with shared ownerships may have similar stock return volatilities, as suggested by some empirical work. See, for example,  \cite{anton2014connected} demonstrate that the degree of shared ownerships is significantly associated with cross-sectional volatility of the stock returns. \cite{li2021information} show that the information transmission between large shareholders has some significant impacts on stock volatility. For this reason, we construct a financial network based on the shared ownerships among these companies as follows.  For each company, we define its {\it major shareholders} as its top 10 shareholders
with more than $1\%$ equity shares.
For a given company $i$, if more than $5\%$ of its total equity shares are held by major shareholders of company $j$, we set $a_{ij}=1$ and otherwise set $a_{ij}=0$. Furthermore, any company that is not connected with other companies is eliminated from the financial network. As a result, we obtain a financial network with $N = 1018$ nodes.  The same type of network structure has been widely used in the literature, e.g.,  \cite{zhu2018network, chen2020community}. However, we wish to comment that other types of the network can be constructed, which can be subsequently  used in the  GNAR model.

\begin{figure}[htpb!]
\centering
\includegraphics[width=0.45\textwidth]{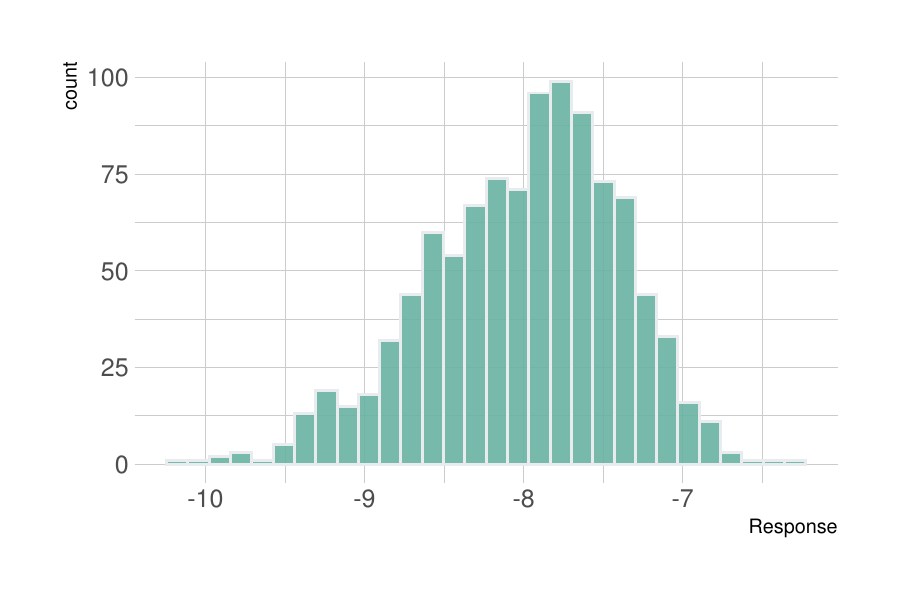}
    \includegraphics[width=0.45\textwidth]{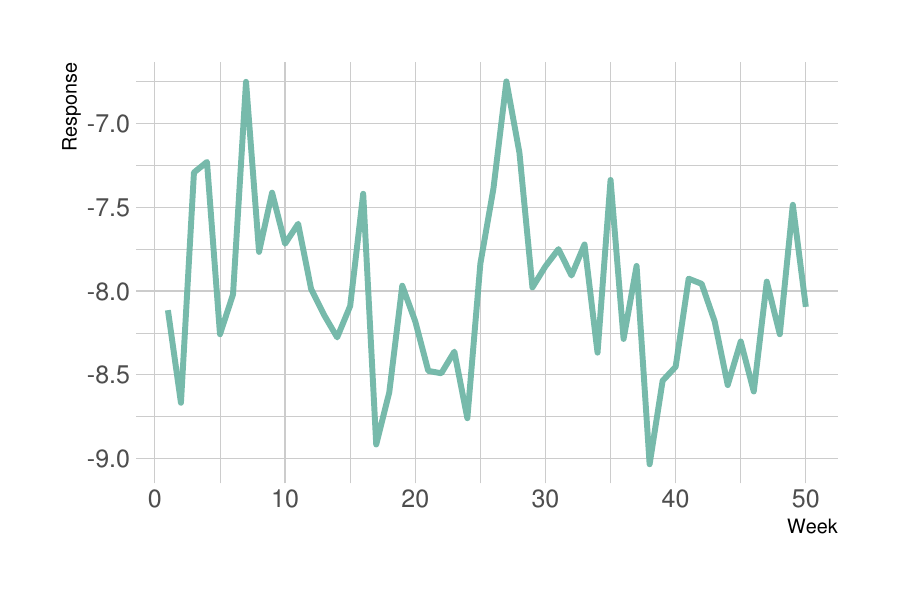}
\caption{\small Left panel: histogram for the temporal averages of the responses (logorithms of  weekly volatilities) for all companies in 2020;
right panel: time series of the crossectional averages of the responses over different companies. }\label{fig:ave}
\end{figure}

For company $i$, we define the response variate $Y_{it}$ as the log-realized weekly return volatility for $T=50$ weeks as following
\[
Y_{it}=\log\left[(K_t-1)^{-1}\sum_{k=2}^{K_t}\left(\log P_{it,k}-\log P_{it,k-1}\right)^2\right],
\]
where $P_{it,k}$ stands for the closing stock price of company $i$ on the $k$th trading day of week $t$, for $t=1,\cdots,T$. A similar measure has been used in~\cite{diebold2014network} to study the network connectivity among financial firms.

The left panel of Figure \ref{fig:ave} presents the histogram of temporal averages of the responses for all $N=1018$ companies and the right panel visualizes the weekly time series on the cross-sectional average of the responses of all companies (i.e., $N^{-1}\sum_i Y_{it}$), where we can observe relatively higher volatility levels during weeks 5--10 and
around the 27th week.
To characterize the dynamic pattern of the stock return volatilities,  motivated by \cite{fama2015five}, we consider the following 6 covariates: SIZE (log-transformed market value), BM (book to market ratio), PR (increased profit ratio compared to the last year),
AR (increased asset ratio compared to the last year),
LEV (log-transformed leverage ratio), and CFM (cash flow divided by market value of the firm).
Lastly, all covariates are  standardized to be mean 0 and variance 1 for later analysis.

\subsubsection{Group Choice and Model Diagnosis}\label{sec:diag}

\begin{figure}[htp]
		\begin{center}
			\subfigure{\includegraphics[trim=1cm 1cm 0.55cm 1cm,clip,scale=0.4]{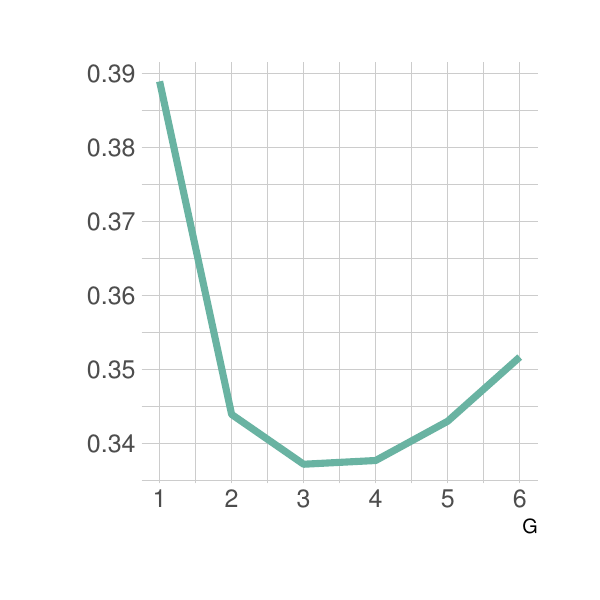}}
		\subfigure{\includegraphics[trim=1cm 1cm 0.55cm 1cm,clip,scale=0.4]{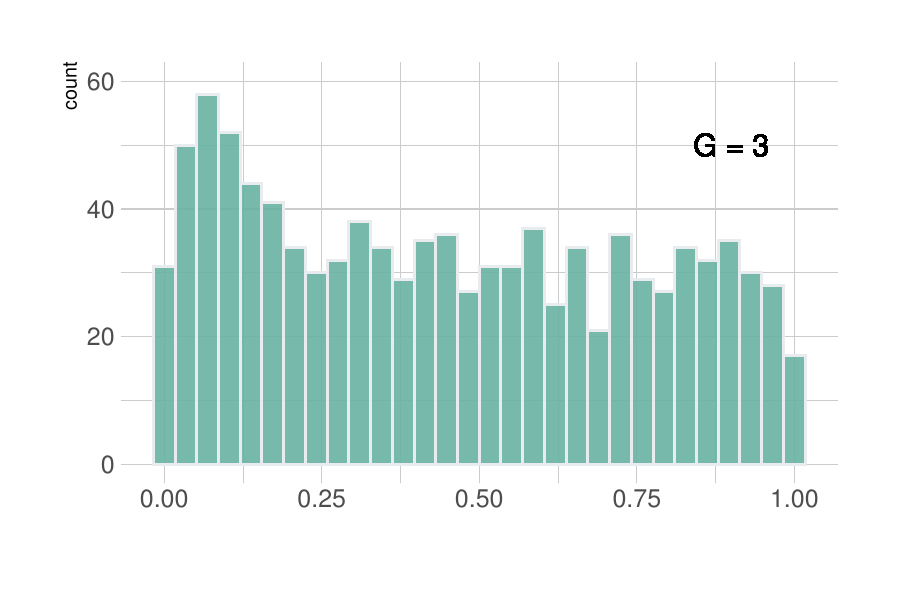}}		
			\subfigure{\includegraphics[trim=1cm 1cm 0.55cm 1cm,clip,scale=0.4]{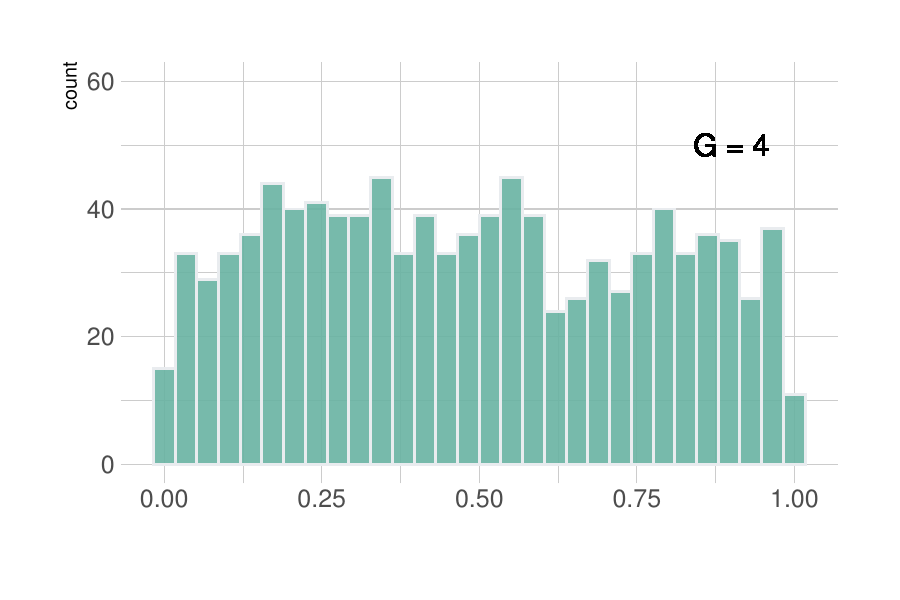}}
			
		\end{center}
		\caption{Left panel: GIC values for $1\le G\le 6$;
				middle and right panels: histograms of $p$-values of node-wise  Ljung-Box test for GNAR models with $ G = 3$ and $4$, respectively.}
		\label{fig:stockG}
	\end{figure}

To apply the GNAR model to the aforementioned dataset, the first task is to choose the number of groups $G$. By setting
$\lambda_{NT} = N^{1/10}T^{-1/2}/(2\min\{10, n_{0.9}\})$ as in the simulation study (recall that $n_{0.9}$ is the 90\% quantile of nodal out-degrees $\{n_i:1\le i\le N\}$), the resulting GIC values indicate that we should select $\wh G = 3$ groups, while $\wh G = 4$ might also be acceptable according to the left panel of Figure \ref{fig:stockG}.

To assess the goodness-of-fit of the GNAR models with $G=3$ or $G=4$, we propose to use the Ljung-Box test \citep{ljung1978measure} to check the serial dependence of the residual time series on each network node. If the GNAR model fits the data sufficiently well, it is expected that the set of residual time series on all network nodes should be close to a set of independent white noise processes. We compute the $p$-values of the Ljung-Box test for a white noise process based on the residual time series collected from each stock and visualize the distribution of $p$-values from all stocks with a histogram, where a large number of small $p$-values may suggest a lack of fit. From the middle (GNAR with $G=3$) and right (GNAR with $G=4$) panels of Figure~\ref{fig:stockG}, we can see that the $p$-values for $G = 4$ are more uniformly distributed than those for $G = 3$, suggesting a better model fit using GNAR with $G=4$.

Deliberating on the GIC score and the diagnostic plots of the model fit, we choose to fit the GNAR model with $G=4$ to the stock data.

\subsubsection{Clustering Results with $\wh G=4$}

\begin{figure}[ht!]
	\begin{center}
					\subfigure{\includegraphics[trim=1cm 1cm 0.55cm 1cm,clip,scale=0.6]{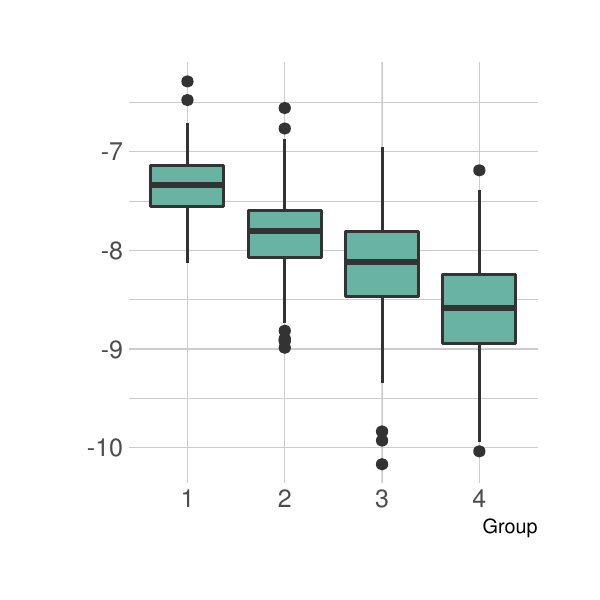}}
					\subfigure{\includegraphics[trim=1cm 1cm 1cm 1cm,clip,scale=0.6]{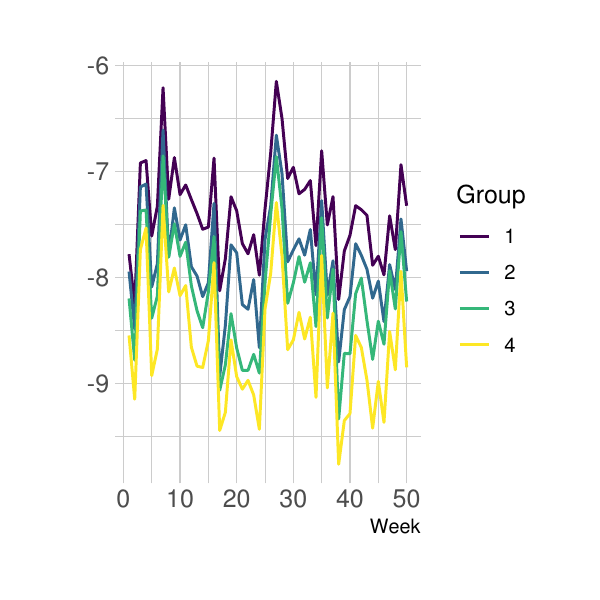}}
	\end{center}
			\caption{Left panel: boxplots of observed average responses over time for $\wh G = 4$ groups; right panel: observed weekly averaged response of $\wh G = 4$ groups over $50$ weeks.}
	\label{fig:stockG2}
\end{figure}

The temporal averages of the responses for different companies are depicted  in the left panel of Figure \ref{fig:stockG2}, where we can see that the first group is of higher volatility levels than
the other 3 groups.
The right panel of Figure \ref{fig:stockG2} visualizes the cross-sectional averages of the responses  within the groups, which shows different dynamic patterns for the four groups.

\begin{figure}[htpb!]
	\centering
	\includegraphics[width=0.8\textwidth]{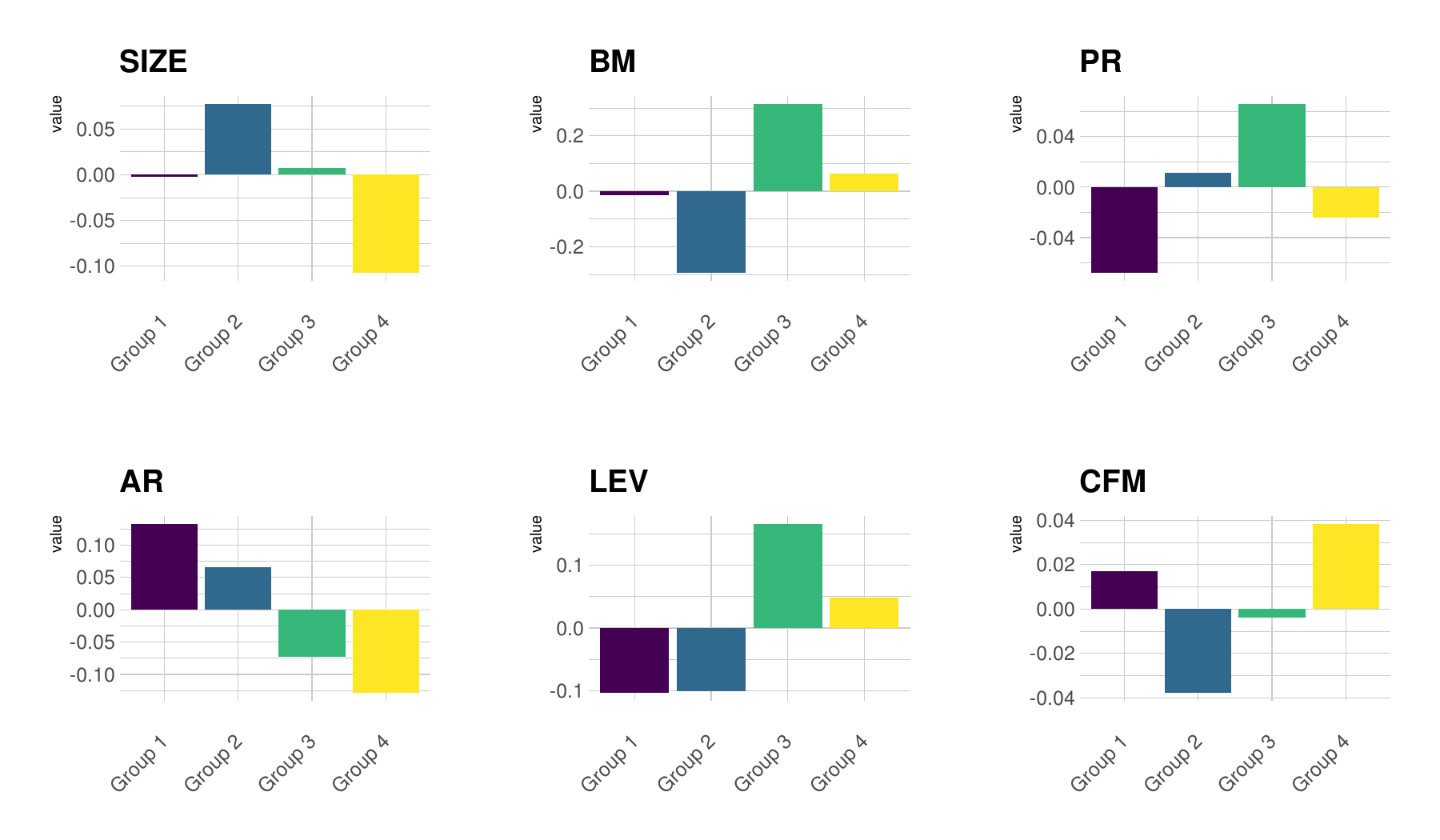}
	\caption{\small Average covariate values within each group. Covariates include:
		SIZE (log-transformed market value), BM (book to market ratio), PR (increased profit ratio compared to the last year),
AR (increased asset ratio compared to the last year),
LEV (log-transformed leverage ratio), and CFM (cash flow divided by market value of the firm).
}\label{fig:ave_covariates}
\end{figure}

To shed more light on the differences among these groups, Figure \ref{fig:ave_covariates} visualizes the average covariate values of different groups.
Specifically, the firms in the second group have the largest size
	while Group 4 has small size firms.
 Group 3 has the largest BM, PR, and LEV values.

\begin{figure}[htpb!]
\centering
\includegraphics[width=0.8\textwidth]{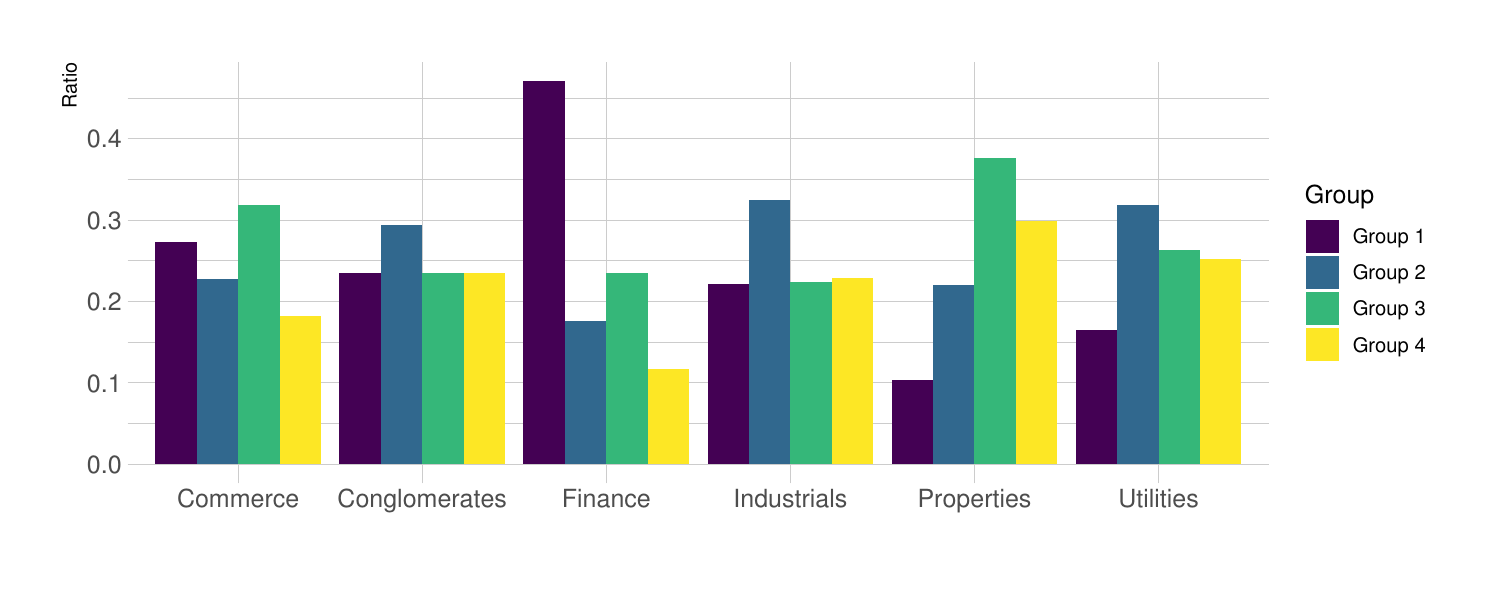}
\caption{\small Proportions of Groups 1, 2 and 3 in six industries (Commerce, Conglomerates, Finance, Industries, Properties, and Utilities).}\label{fig:industry}
\end{figure}

Lastly, we summarize the industry information of companies in each group in Figure
\ref{fig:industry}.
These companies can be roughly categorized into six major industries (Commerce, Conglomerates, Finance, Industries, Properties, and Utilities) according to the information released by China Securities Regulation Commission (CSRC) in 2012.
We can see that most companies in the Finance  are clustered in Group 1. Most companies in industrials and Utilities are clustered into Group 2 while most companies in Properties are in Group 3.
Note that companies in the Industrials category typically have large market values, which explains why the averaged SIZE is the highest for Group 2 in Figure~\ref{fig:ave_covariates}.

\subsubsection{Model Interpretations}

The model estimation results are given in Table \ref{reg2}.
First, we observe that the momentum effect  is the highest in Group 3
and the lowest in Group 2.
This suggests that stock return volatilities of companies in Group 3 are
highly influenced by its historical performance.
Furthermore, the within-group network effects of all groups are rather similar and positive except Group 3, indicating a volatility spillover effect within each group.
Regarding the between-group network effects, we can see that
Groups 1, 2 and 4 positively influence each other with significant between-group network effects. In contrast,
Group 3 receives negative influences from the other groups.

	\begin{table}[htp]
	\caption{The model parameter estimates and associated $p$-values for the stock data, where
		``*" denotes statistical significance at the
		0.05 level. }\label{reg2}
	\begin{center}
		\scalebox{0.85}{
			\begin{tabular}{l|rrrr}
				\hline
				\hline
				& {\sc Group 1 } &{\sc Group 2 } &{\sc Group 3 } &{\sc Group 4 }  \\
				\hline
				{\sc Proportion} &  0.215 & 0.302 & 0.250 & 0.234 \\
{\sc Group 1 } & 0.095 * ($<0.001$) & 0.082 * ($<0.001$) & 0.100 * ($<0.001$) & 0.072 * ($<0.001$) \\
{\sc Group 2 } & 0.054 * ($<0.001$) & 0.069 * ($<0.001$) & 0.080 * ($<0.001$) & 0.077 * ($<0.001$) \\
{\sc Group 3 } & -0.075 * ($<0.001$) & -0.093 * ($<0.001$) & -0.022  (0.052) & -0.064 * ($<0.001$) \\
{\sc Group 4 } & 0.047 * ($<0.001$) & 0.035 * (0.005) & 0.047 * ($<0.001$) & 0.066 * ($<0.001$) \\
{\sc Momentum } & 0.242 * ($<0.001$) & 0.089 * ($<0.001$) & 0.433 * ($<0.001$) & 0.192 * ($<0.001$) \\
\hline
{\sc Intercept } & -4.854 * ($<0.001$) & -6.659 * ($<0.001$) & -5.115 * ($<0.001$) & -6.573 * ($<0.001$) \\
{\sc SIZE } & -0.107 * ($<0.001$) & -0.123 * ($<0.001$) & -0.093 * ($<0.001$) & -0.136 * ($<0.001$) \\
{\sc BM } & -0.108 * ($<0.001$) & -0.247 * ($<0.001$) & -0.118 * ($<0.001$) & -0.252 * ($<0.001$) \\
{\sc PR } & -0.015  (0.259) & 0.024 * (0.021) & 0.018 * (0.023) & 0.069 * ($<0.001$) \\
{\sc AR } & -0.028 * (0.009) & -0.039 * ($<0.001$) & -0.059 * ($<0.001$) & -0.034 * (0.023) \\
{\sc LEV } & 0.070 * ($<0.001$) & 0.079 * ($<0.001$) & 0.089 * ($<0.001$) & 0.078 * ($<0.001$) \\
{\sc CFM } & -0.062 * ($<0.001$) & -0.099 * ($<0.001$) & -0.057 * ($<0.001$) & -0.072 * ($<0.001$) \\
				\hline
				\hline
			\end{tabular}
		}
	\end{center}
\end{table}

Finally, we comment on the estimated coefficients of the covariates.
First, the volatility levels have negative relationships with the
market values (SIZE) of the firms across all groups.
This confirms the phenomenon that firms with larger sizes tend to perform better when exposed to financial risk than smaller firms \citep{diebold2014network,huang2021network}.
The BM, AR and CFM value are also shown to have significant
negative effects on the volatility level across all groups.
On the contrary, the LEV tends to have a positive effect on the volatilities and the PR values are also shown
to have a positive influence on most groups.


\subsection{Model Prediction}

Lastly, we compare the prediction performance of the GNAR model with the LS-VAR model \citep{basu2019low}.
Specifically, we use the first $T_{tr} = 40$ weeks for model training and the following $T_{test} = 10$ weeks for model testing.
For the GNAR model, we use $\wh G = 4$ groups.
For the LS-VAR model, we choose the tuning parameters by minimizing the prediction RMSE on the testing dataset, which is defined as $\mbox{PRMSE} = \{\sum_{i  =1}^N\sum_{t = T_{tr}+1}^T (\wh Y_{it} - Y_{it})^2/(NT_{test})\}^{1/2}$. Since the LS-VAR does not allow time-invariant covariates used in the GNAR model, we center the observations by $\wt Y_{it}=Y_{it}-\ol Y_i$ with $\ol Y_i=\frac{1}{T}\sum_{t=1}^{T_{tr}}Y_{it}$ for $i=1,\cdots, N$. As a result,
the PRMSE values for the GNAR model and LS-VAR model are 1.16 and 1.20, respectively.
Although the difference is relatively small, the proposed GNAR model is able to achieve slightly lower prediction error with fewer model parameters.


\subsection{User Activity Analysis with Sina Weibo}

	\label{sec:weibo}
	In this section, we illustrate the use of the proposed methodology with a dataset collected from Sina Weibo, a Twitter-type online social network platform in China.
	The dataset includes $N = 804$ active users, whose posting activities are recorded for $T = 75$ days. The network structure is obtained using the observed following-followee relationships among the users.
	
	To gauge the users' activity levels, we follow \cite{zhu2017network} to define the response $\wt Y_{it}=\log(1+X_{it})$ with $X_{it}$ being the number of posts of the $i$th user in the $t$th day.
	To remove the time-varying trend, we center the response variable as $ Y_{it} = \wt Y_{it} - N^{-1}\sum_{j=1}^N\wt Y_{jt}$. To further explain the variations of $Y_{it}$'s among different network nodes, we collect seven node-specific covariates: {\sc Gender} (male = 1, female = 0),
	{\sc Tenure} (number of years since the user's registration),
	{\sc Beijing} (equals to 1 if the user locates in Beijing and 0 otherwise),
	{\sc Shanghai} (equals to 1 if the user locates in Shanghai and 0 otherwise),
	{\sc Description} (the length of user self-description),
	{\sc Weibo} (logarithm of accumulated number of posts),
	and {\sc Public} (equals to 1 if the user is a public account and 0 otherwise).
	
\subsubsection{Group Choice and Clustering Results}
	
\begin{figure}[htp]
		\begin{center}
			\subfigure{\includegraphics[trim=1cm 1cm 0.55cm 1cm,clip,scale=0.6]{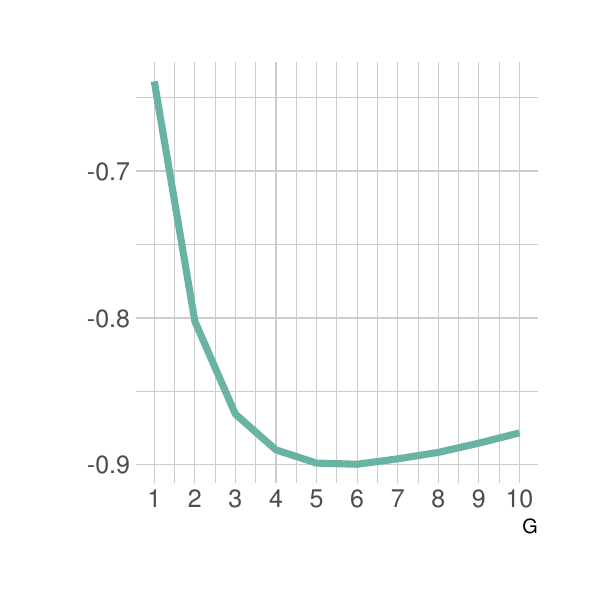}}
			\subfigure{\includegraphics[trim=1cm 1cm 0.55cm 1cm,clip,scale=0.6]{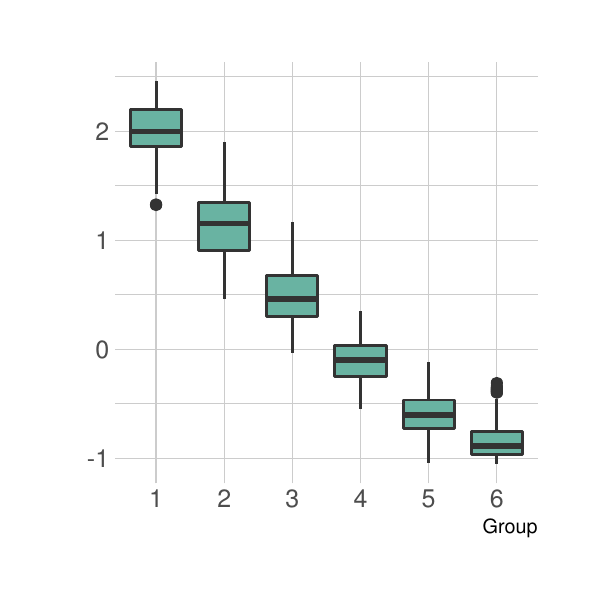}}
			\subfigure{\includegraphics[trim=1cm 1cm 1cm 1cm,clip,scale=0.6]{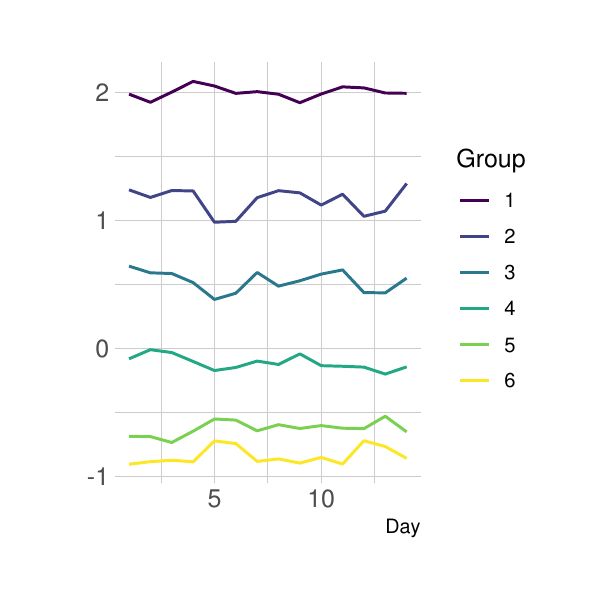}}
		\end{center}
		\caption{Left panel: GIC values for $1\le G\le 10$ with the minimum at $\wh G = 6$;
			middle panel: boxplots of observed average user response over time for $\wh G = 6$ groups; right panel: observed average daily response of $\wh G = 6$ groups over two consecutive weeks.}
		\label{fig:weiboG}
	\end{figure}

	We start by choosing the number of groups $G$ using the GIC criterion defined in~\eqref{qic} with a 	$\lambda_{NT} = N^{1/10}T^{-1/2}/(2\min\{10, n_{0.9}\})$,
	where $n_{0.9}$ is the 90\% quantile of nodal out-degrees $\{n_i:1\le i\le N\}$.
	As illustrated in the left panel of Figure~\ref{fig:weiboG},  the GIC value is minimized at $\wh G=6$, although  $\wh G=5$ is also acceptable. One notable feature is that the GIC achieves significant reductions by increasing from $G=1$ to $G=4$, suggesting that there indeed exists certain level heterogeneity among network users. This provides some justifications for the proposed method that introduce latent groups among the network nodes.

\subsubsection{Model Interpretations}
	
	Next, we obtain the parameter estimates and group membership assignments by minimizing~\eqref{Q_obj1}, which are summarized in Table~\ref{reg}  and Figure~\ref{fig:weiboG}. In Figure~\ref{fig:weiboG}, The middle panel indicates clear different individual activity levels for users in different groups, and the right panel reveals rather consistent separations for overall group activity  level over time. In particular, Groups 2 and 3 tend to be less active during weekends while the other three groups do not exhibit such a pattern.

	\begin{table}[htp]
		\caption{The model parameter estimates and associated $p$-values for the Sina Weibo data, where
			``*" denotes statistical significance at the
			0.05 level.}\label{reg}
		\begin{center}
			\scalebox{0.63}{
				\begin{tabular}{l|rrrrrr}
					\hline
					\hline
					& {\sc Group 1 } &{\sc Group 2 } &{\sc Group 3 }   &{\sc Group 4 } &{\sc Group 5 } &{\sc Group 6 }\\
					\hline
					{\sc Proportion} &  0.086 & 0.109 & 0.134 & 0.200 & 0.221 & 0.249 \\
					{\sc Group 1 } & 0.332 * ($<0.001$) & 0.042  (0.691) & 0.150  (0.316) & -0.231  (0.151) & 0.215  (0.185) & -0.247  (0.053) \\
					{\sc Group 2 } & 0.492 * ($<0.001$) & 0.746 * ($<0.001$) & 0.643 * ($<0.001$) & -0.226  (0.257) & 0.506 * (0.009) & 0.105  (0.520) \\
					{\sc Group 3 } & 0.453 * ($<0.001$) & 0.197  (0.132) & 0.190  (0.318) & -0.312  (0.091) & -0.154  (0.370) & -0.103  (0.467) \\
					{\sc Group 4 } & 0.194 * ($<0.001$) & 0.290 * (0.018) & 0.106  (0.496) & -0.155  (0.337) & 0.187  (0.181) & -0.061  (0.605) \\
					{\sc Group 5 } & -0.090 * (0.047) & 0.053  (0.597) & -0.167  (0.201) & 0.454 * ($<0.001$) & 0.050  (0.669) & 0.119  (0.262) \\
					{\sc Group 6 } & -0.062  (0.112) & -0.044  (0.575) & -0.301 * (0.005) & 0.289 * (0.010) & 0.286 * (0.003) & 0.200 * (0.013) \\
					{\sc Momentum } & 0.523 * ($<0.001$) & 0.293 * ($<0.001$) & 0.338 * ($<0.001$) & 0.286 * ($<0.001$) & 0.234 * ($<0.001$) & 0.157 * ($<0.001$) \\
					\hline
					{\sc Intercept } & 0.731 * ($<0.001$) & -0.044  (0.549) & -0.288 * ($<0.001$) & -0.563 * ($<0.001$) & -0.606 * ($<0.001$) & -0.812 * ($<0.001$) \\
					{\sc Gender } & 0.021  (0.126) & 0.041 * (0.015) & -0.034 * (0.048) & 0.037 * (0.008) & 0.031 * (0.011) & 0.016  (0.116) \\
					{\sc Tenure } & 0.023 * (0.015) & 0.106 * ($<0.001$) & 0.067 * ($<0.001$) & 0.057 * ($<0.001$) & 0.060 * ($<0.001$) & 0.025 * ($<0.001$) \\
					{\sc Beijing } & 0.037 * (0.009) & 0.201 * ($<0.001$) & 0.086 * ($<0.001$) & 0.015  (0.403) & -0.038 * (0.013) & -0.026 * (0.016) \\
					{\sc Shanghai } & -0.005  (0.824) & 0.259 * ($<0.001$) & 0.114 * ($<0.001$) & 0.076 * ($<0.001$) & -0.286 * ($<0.001$) & 0.284 * ($<0.001$) \\
					{\sc Description } & -0.011  (0.054) & -0.013 * (0.024) & 0.018 * (0.003) & 0.028 * ($<0.001$) & 0.032 * ($<0.001$) & 0.006  (0.081) \\
					{\sc Weibo } & -0.004  (0.244) & 0.013 * (0.005) & 0.010  (0.052) & 0.010 * (0.020) & -0.004  (0.243) & 0.010 * ($<0.001$) \\
					{\sc Public } & 0.046 * (0.023) & 0.135 * ($<0.001$) & 0.066 * (0.009) & 0.135 * ($<0.001$) & 0.027  (0.092) & 0.005  (0.662) \\
					\hline
					\hline
				\end{tabular}
			}
		\end{center}
	\end{table}


	From Table~\ref{reg},
	we can observe some interesting dynamic patterns among the six groups.
	Firstly, Group 1 appears to be the most self-excited group who has the largest momentum effect (i.e. 0.523). In the meantime, Group 1 also appears to be the most influential group in the sense that 5 out of $\wh \beta_{g1}$, $g=1,\cdots,6$ are statistically significant, suggesting users in other groups tend to be influenced by users in Group 1.
	Secondly, Group 2 has the largest within-group network effect (i.e., 0.746) but has little impact on activities of other groups except for Group 4. Activities of users in Group 2 are also heavily influenced by activities of other groups. For example, Group 2 receives significant positive network influence from Group 3 but
	its impact on Group 3 is not significant, which implies an asymmetric influential pattern.
	Lastly,
	the activities of Group 6 appear to be
	positively related to Groups 4--5, but not to the most influential Group 1.
	
	For the fixed-effects,	we observe that the male users tend to be more active in Groups 2, 4, 5 but less active in Group 3.
	The  users with longer tenure tend to be more active in all groups.
	For the location related covariates,
	we observe that the users located in Beijing of Groups 1--3
	are more active while users in Shanghai of Groups 2, 3, 4, 6 tend to be more active.
	Lastly, the activity levels of Groups 3, 4 and 6
	are positively related to their historical accumulated Weibo posts and
	the public accounts tend to be more active in Groups 1--4.

\subsubsection{Model Diagnosis and Improvement}
\begin{figure}[htp]
		\begin{center}
			\subfigure{\includegraphics[trim=1cm 1cm 0.55cm 1cm,clip,scale=0.32]{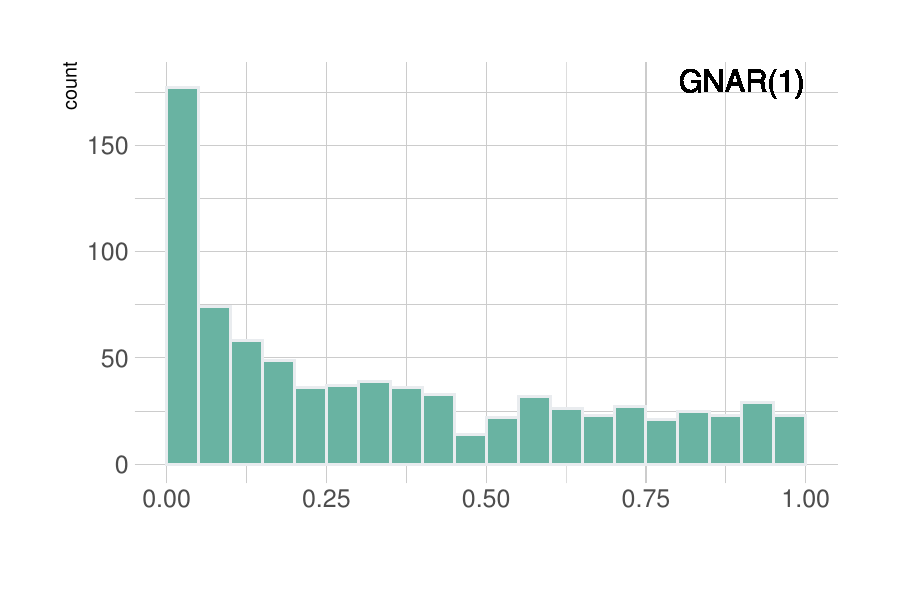}}
		\subfigure{\includegraphics[trim=1cm 1cm 0.55cm 1cm,clip,scale=0.32]{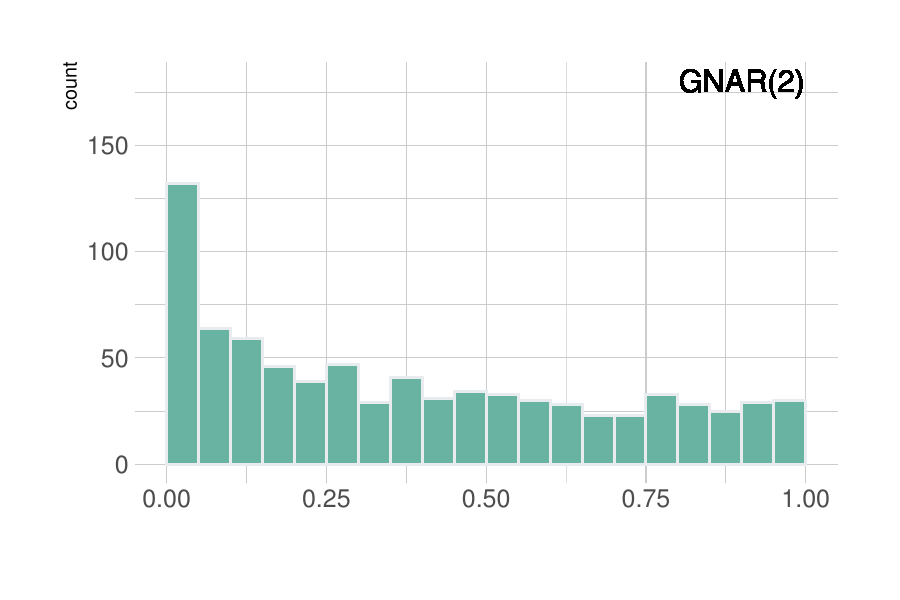}}		
			\subfigure{\includegraphics[trim=1cm 1cm 0.55cm 1cm,clip,scale=0.32]{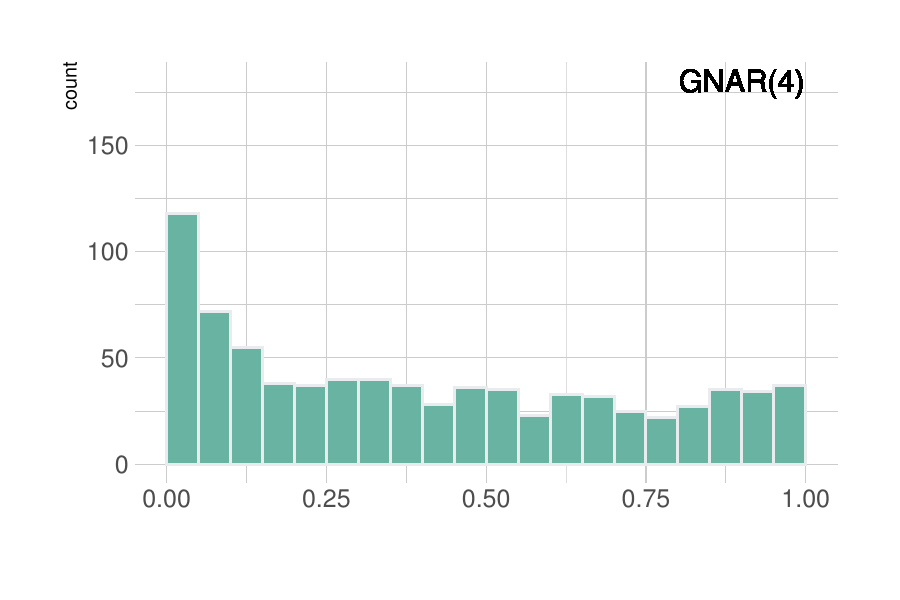}}
			
		\end{center}
		\caption{Histograms of $p$-values of Ljung-Box tests for the GNAR model (left), the   GNAR(2) model (middle), and the GNAR(4) model (right), all of which use $\wh G=6$.}
		\label{fig:weiboG2}
	\end{figure}

	Following the same idea in Section~\ref{sec:diag}, we use the histograms of the $p$-values of the Ljung-Box tests to access the goodness of fit. From the left panel of Figure~\ref{fig:weiboG2}, we can see that the distribution of $p$-values is far from a uniform distribution, suggesting a lack of fit with the proposed GNAR model to the Weibo data. Therefore, the model interpretations given in sections 5.2.1-5.2.2 should be treated with caution.

	In our further attempts to improve the model fit, we implemented the following GNAR($q$) model as a direct extension of the proposed GNAR model~\eqref{gnar},
	\begin{align}
		Y_{it} =  \sum_{k=1}^{q}\sum_{j = 1,j\ne i}^N\beta_{g_ig_j,k}w_{ij}Y_{j(t-k)} + \sum_{k=1}^{q}\nu_{g_i,k} Y_{i(t-k)} +
		\bz_i^\top\bzeta_{g_i} +
		\ve_{it},\quad t=1,\cdots,T.\label{gnarp}
	\end{align}
	It is straightforward to see that when $q=1$, the above model reduces to  model~\eqref{gnar}. We apply the  GNAR(2) and the  GNAR(4) to the Weibo data, whose diagnostic plots are given in Figure~\ref{fig:weiboG2}. We can observe that, by increasing $q$ from $1$ to $4$, the diagnostic plot indeed becomes closer to a uniform distribution but fails to fully address the lack-of-fit issue. To further improve the model, more relevant covariates including some time-dependent covariates may be collected, which will be pursued in a separate work.

	\section{Concluding Remarks}

	In this work, we propose a network vector autoregression model with a latent group structure.
	The flexibility of the model enables us to capture the individuals' heterogeneous momentum effects and network interactions.
	Group memberships and model parameters are estimated simultaneously through the minimization of a least-square type loss function, and the theoretical properties of the resulting estimators are investigated. Furthermore, a data-driven criterion is designed to consistently select the number of groups. The usefulness of the proposed model is illustrated through simulation studies and two real data examples.
	
	To conclude the article, we discuss several interesting future research topics. First, one immediate extension of the current work is to investigate theoretical properties of the GNAR($q$) model suggested in~\eqref{gnarp}, including the asymptotic normality, choice of $q$, and the goodness-of-fit tests. Second, the group structure in the proposed model is primarily determined by node-specific characteristics but not by interactions among different nodes. It will be interesting to combine the proposed group structure with some community detection methods for a more practical model.
Third, the fixed effects are assumed to be parametric and time-invariant. It is desirable to extend the current setting to include nonparametric and/or time-varying fixed effects.
	Lastly, the covariates considered in our work are of finite dimension. However, in practice high dimensional features can be collected. Feature screening and selection techniques can be developed to uncover the most informative features.

	\baselineskip = 5 mm
	\bibliographystyle{asa}
	\bibliography{xuening}
	
		\baselineskip = 7 mm
	\section*{Appendix: Initial Membership Estimation}
	\renewcommand{\theequation}{A.\arabic{equation}}
	In this section, we propose a $k$-means type algorithm to obtain an initial membership estimator $\wh \mG^{(0)}$. Define $\ol Y_i=T^{-1}\sum_{t=1}^TY_{it}$, $\ol Y_{i,lag}=T^{-1}\sum_{t=0}^{T-1}Y_{it}$, and correspondingly  $\wt Y_{it} = Y_{it} - \ol Y_i$ and $\wt Y_{it,lag} = Y_{it} - \ol Y_{i,lag}$. Then based on model~\eqref{gnar}, one has that
	\be
	\label{eq:a}
	\begin{split}
		\wt Y_{it}&=\sum_{j=1}^N \beta_{g_ig_j}w_{ij}\wt Y_{j(t-1),lag}+\nu_{g_i}\wt Y_{i(t-1),lag}+\wt\varepsilon_{it},\\
		\ol Y_i&=\sum_{j=1}^N \beta_{g_ig_j}w_{ij}\ol Y_{j,lag}+\nu_{g_i}\ol Y_{i,lag}+\bz_i^\top\bzeta_{g_i}+\ol\varepsilon_i,
	\end{split}
	\ee
	where $\ol\varepsilon_i=T^{-1}\sum_{t=1}^T\varepsilon_{it}$ and $\wt\varepsilon_{it}=\varepsilon_{it}-\ol\varepsilon_i$, $i=1,\dots,N$, $t=1,\dots,T$. The first equation removes the fixed heterogeneous effect through centering, from which network effect $\beta$ and momentum effect $\nu$ can be estimated by treating each node as a group. This gives an crude but unbiased initiate estimates (if there are sufficient data so that the least-squares can be used).  With estimated parameter, the second equation gives an estimate of the fixed effect.
	
	To make the above idea more precise,  let $\bx_{it} = (( w_{ij}\wt Y_{j(t-1),lag}: j\in \mN_i)^\top, \wt Y_{i(t-1),lag})^\top\in \mR^{n_i+1}$, where $\mN_i=\{j: a_{ij}\ne 0\}$.  Then based only on observations from node $i$, we obtain the following two estimates from \eqref{eq:a}:
	\begin{align*}
	\wh \bb_{i} &= (\wh b_{i1},\wh b_{i2},\cdots, \wh b_{in_i},\wh v_i )^\top = \Big(\sum_{t=1}^T \bx_{it}\bx_{it}^\top + \lambda \bI_{n_i+1}\Big)^{-1}\Big(\sum_{t=1}^T \bx_{it}\wt Y_{it}\Big),\\
	\wh f_{i} &= \wh{\bz_i^\top\bzeta}=\ol Y_i-\sum_{j=1}^N \wh b_{ij}w_{ij}\ol Y_{j,lag}-\wh v_i\ol Y_{i,lag},
	\end{align*}
	where $\lambda = 0.01\times \sum_t\|\bx_{it}\|^2/(n_i+1)+10^{-6}$ is a ridge tuning parameter. We use the following three $k$-means algorithms to obtain multiple initial membership  vector $\wh \mG^{(0)}$'s.
	\begin{enumerate}
		\item $k$-means based on individual momentum parameter estimates $\wh v_1,\cdots,\wh v_N$.
		\item $k$-means based on individual fixed-effect estimates  $\wh f_1,\cdots,\wh f_N$.
		\item $k$-means based on individual network effect estimates $\wh b_{ij}$'s for $j=1,\cdots,n_i$, $i=1,\cdots,N$, using following steps.
		\begin{itemize}
			\item [(1)] Run a $k$-means algorithm over the collection of estimated network effects $\{\wh b_{ij}: j=1,\cdots,n_i,i=1,\cdots,N \}$ with the number of clusters as $k = G^2$. The cluster label of $\wh b_{ij}$ is denoted as $c_{ij}\in [G^2]$.
			
			\item [(3)]
			For each node $i$, define a $G^2\times 1$ vector $\wt \bb_{(i)}^{net} = (\wt b_{i1},\cdots,\wt b_{iG^2} )^\top$ where we define $\wt b_{il}=({\sum_j \wh b_{ij}I(c_{ij} = l)})/({\sum_{j = 1}^{n_i}I(c_{ij} = l)})$ for $1\le l\le G^2$. Next, define $\wt \bb_{(i)} = (\wh v_i, \wt \bb_{(i)}^{net\top})^\top$, and run a $k$-means algorithm over set $\wt \bb_{(1)},\cdots,\wt \bb_{(N)}$ with a $k=G$ groups. The resulting membership vector is a possible value for $ \wh \mG^{(0)}$.
		\end{itemize}
	\end{enumerate}
	The intuition behind the third $k$-means algorithm is as follows. There are at most $G^2$ distinct values in the network parameter vector $\bbeta$, hence we first cluster $\wh b_{ij}$'s into $G^2$ groups. Then by the definition of  $\wt \bb_{(i)}$'s, if nodes $i,j$ belong to the same group, one can expect that $\wt \bb_{(i)}\approx \wt \bb_{(j)}$. Therefore, we can apply the $k$-means algorithm to $\wt \bb_{(i)}$'s for
	an initial estimate $\wh \mG^{(0)}$.
	
	In our numerical examples, we repeat the above three $k$-means algorithms for $100$ times with different initialization seeds and use the resulting $\wh \mG^{(0)}$'s for the minimization of the proposed algorithm for~\eqref{Q_obj1}.

\end{document}